# Physical Regimes of Electrostatic Wave-Wave nonlinear interactions generated by an Electron Beam Propagating in a Background Plasma

Haomin Sun[1,2,4*], Jian Chen[3+], Igor D. Kaganovich[1], Alexander Khrabrov[1], Dmytro Sydorenko[5]

[1]Princeton Plasma Physics Laboratory, Princeton University, Princeton, New Jersey 08543, USA

[2]CAS Key Laboratory of Geospace Environment, Department of Geophysics and Planetary Science, University of Science and Technology of China, Hefei, Anhui, People's Republic of China

[3]Sino-French Institute of Nuclear Engineering and Technology, Sun Yat-sen University, Zhuhai 519082, P. R. China

[4]CAS Center for Excellence in Comparative Planetology, People's Republic of China

[3]University of Alberta, Edmonton, Alberta T6G 2E1, Canada

Corresponding Authors: Haomin Sun, Email: haomins@princeton.edu
                      Jian Chen, Email: chenjian5@mail.sysu.edu.cn

# Abstract

　Electron-beam plasma interaction has long been a topic of great interest. Despite the success of Quasi-Linear (QL) theory and Weak Turbulence (WT) theory, their validities are limited by the requirement of sufficiently dense mode spectrum and small wave amplitude. In this paper, we extensively studied the collective processes of a mono-energetic electron beam emitted from a thermionic cathode propagating through a cold plasma by performing a large number of high resolution two-dimensional (2D) particle-in-cell (PIC) simulations and using analytical theories. We confirm that during initial stage of two-stream instability between the beam and background cold electrons it is saturated due to well-known wave-trapping mechanism. Further evolution occurs due to strong wave-wave nonlinear processes. We show that the beam-plasma interaction can be classified into four different physical regimes in the parameter space for the plasma and beam parameters. The differences between the different regimes are analyzed in detail. For the first time, we identified a new regime in strong Langmuir turbulence featured by what we call Electron Modulational Instability (EMI) that could create a local Langmuir wave packet growing faster than the ion plasma frequency. Ions do not have time to respond to EMI in the initial growing stage. On a longer timescale, the action of the ponderomotive force produces very strong ion density perturbations, and eventually the beam-plasma wave interaction stops being resonant due to strong ion density perturbations. Consequently, in this EMI regime, electron beam-plasma interaction occurs in a periodic (intermittent) process. The beams are strongly scattered by waves, and the Langmuir wave spectrum is significantly broadened, which in turn gives rise to strong heating of bulk electrons. Associated energy transfer from the beam to the background plasma electrons has been studied. A resulting kappa distribution and a wave-energy spectrum, $E^2(k) \sim k^{-5}$, are observed in the strong turbulent regime.

# 1. Introduction

Electron beam-plasma interaction has long been a topic of great interest yielding many results relevant to the nonlinear Langmuir waves and turbulence [1-5]. It is also believed to be an ubiquitous phenomena in many areas such as beam-generated plasma discharges [6-8] (where an electron beam is typically injected from a source or a biased electrode with emission), space plasmas [9-14], in particularly in regard to formation of phase space holes [15-18]. A number of theories have been proposed to describe the electron beam-plasma interaction. The linear theory was first developed by Landau [19], Akhiezer et al. [20], and Bohm & Gross [21], providing the linear growth rate of the two-stream instability. To further consider the wave-particle coupling, quasi-linear (QL) theory [22,23] was then developed, which described the evolution of the electron velocity distribution function (EVDF) as the diffusion in velocity space and corresponding wave spectrum of the electric field using the linear plasma dispersion relation. Subsequently, a weak turbulence (WT) theory (improved based on QL theory) was proposed by S. A. Kaplan and V. N. Tsytovich [24] to take into account wave-wave interactions. Recently, this theory was further advanced by P. Yoon [25,26]. The WT theory considers the wave-wave coupling and the wave-particle resonances and the nonlinear scattering of electrons and ions on different waves [27]. Based on this theory, the self-consistent generation of super-thermal electrons [28] and the emission of electromagnetic (EM) waves can be theoretically described [29]. However, both the QL and WT theories are only applicable for the cases with a sufficiently weak beam or small-amplitude waves [1,30].

For the cases where an intense mono-energetic beam is injected into a plasma, a strong Langmuir turbulence can be excited, in which the nonlinear wave-wave interaction becomes dominant; and where a modulational wave-wave instability comes into play and begins to govern the nonlinear evolution of the Langmuir waves, calling for a more appropriate theory incorporating these modulational effects. V.E. Zakharov [31] first proposed the Langmuir collapse theory based on the now well-known Zakharov equations to interpret the growth of Langmuir wave packets in strong Langmuir turbulence (SLT). The theory showed that the Langmuir wave packets undergo the self-similar focusing process allowing the energy transfer to smaller scales, during which the quasi-neutral plasma is driven by the ponderomotive force of an increasing electric field, thereby leading to large density perturbations. This phenomenon is known as Langmuir collapse. Since then, many follow-up studies were carried out, including the simulations of the strong turbulence by numerically solving the Zakharov equations [32-35] and the observations of electron heating in ionosphere plasmas due to Langmuir collapse [12,36,37].

Despite the success of the Zakharov equations in describing the nonlinear evolution of the Langmuir wave packets, these equations only rigorously describe the wave-wave instabilities on the time scale comparable to the ion plasma period ($\sim 1/\omega_{pi}$) because a quasi-neutral assumption was used (see more details in Appendix 1). *Therefore, the faster process on the time scale between $1/\omega_{pi}$ and $1/\omega_{pe}$ could not be modeled self-consistently using the Zakharov equations.* Such a faster process has been observed in the beam-plasma laboratory experiments [38], showing that the Langmuir wave packets can grow faster than the ion response. The exact physical mechanism for this fast process was unclear. *Understanding the mechanisms for the different regimes of the wave-wave instabilities created by an electron beam and their respective onset thresholds are the focus of the present study.*

In addition, we also take into consideration the kinetic effects or wave-particle interactions due to the importance of super-thermal electrons for the nonlinear damping to the waves, which could be very important in the late stage of Langmuir collapse [39]. Such effects are harder to include self-consistently into the Zakharov equations, despite several transit-time damping models [40-43] have been proposed to try to mitigate this problem. Therefore, for a more accurate understanding of the nonlinear beam-plasma interaction, a comprehensive kinetic simulation study was performed in this paper, using the well-developed particle-in-cell (PIC) code EDIPIC-2D [44].

We extensively studied the nonlinear beam-plasma interaction in a DC discharge with an emissive cathode, where an intense mono-energetic electron beam is generated by the electron emission from the cathode and the subsequent acceleration in a cathode sheath [6-8]. Our primary focus in this study is the nonlinear physics near the electron beam injection point, which is relevant to many low temperature plasma devices that use the electron beam-generated plasmas [6,45-47] including ubiquitous hollow cathodes [48]. In those applications, there is need to determine how quickly beam heats cold electrons for different experimental conditions as well as the physical processes in the systems. We proposed classification of different instability regimes that is relevant not only to low temperature plasmas but also has a much wider range of applications. We identified four different regimes characterized by different wave-wave nonlinear instabilities. The details of nonlinear wave coupling, such as the energy transfer, energy spectrum, and beam scattering, are presented. Most importantly, for the first time, we identified a new modulational instability in strong Langmuir turbulence (we has named it the *electron modulation instability,* EMI) which has different characteristics from the commonly known modulational (MI) and parametric-decay instabilities (PDI). An analytical theory was developed to clarify the physical mechanism governing the generation of SLT and in particularly in EMI regime. The proposed analytical criteria for different regimes were verified by analyzing 57 simulation cases. The present work provides detailed information on the observed SLT properties, which can also be used for further development of the turbulence theory. This paper is a joint submission of a short letter in *Physical Review Letters* [49], where we focus on the physical properties of EMI regime.

This paper is organized as follows: The simulation model is thoroughly introduced in Section 2. Simulation results and discussions regarding different physical regimes of the nonlinear beam-plasma interaction are presented in Section 3. The theory for strong turbulence and EMI is described in Section 4, and a summary is provided in Section 5. Details of the derivations of the proposed theory are given in Appendix.

## 2. Simulation Model

For the present study, we adopted a two-dimensional slab calculation domain ($L_x \times L_y$) consisting of two flat electrodes and an argon plasma (see Fig. 1). The cathode located at $x = 0$ is biased at a constant negative value $V_0$, and the anode at $x = L_x$ is grounded. Periodic boundary conditions are applied to the other two boundaries ($y = 0$ and $y = L_y$). Here, we use dimensional units for all the physical quantities. To help readers better understand the physical mechanisms, a transformation between dimensional and dimensionless units is given in Table 1. The initial plasma parameters are set as follows, the uniform plasma density is $n_{p0} = n_{e0} = n_{i0} = 10^{17} m^{-3}$, the ion temperature, $T_{i0} = 0.03 eV$, and the bulk electron temperature, $T_{e0}$ was varied in a large parameter range. The cell size, $\Delta x = \Delta y = 0.0083 mm$, and time step, $\Delta t = \Delta x/v_{max}$ are used, where $v_{max}$ is the maximum electron velocity of particles expected in simulations. In all the simulation cases, constant value of $L_y = 9mm$ was used, while $L_x$ varied. Initially, 800 macro-particles per cell for each species are created. An electron beam with varied density $n_b/n_{p0} = 0.00068 - 0.07$ and constant beam temperature $T_{eb} = 0.2eV$ is uniformly injected from the cathode after $t = 80ns$. Here, an initial delay of $t \approx 5\omega_{pi}^{-1} \approx 80ns$ was used for all the 57 simulation cases in order to form a steady-state sheath and therefore to avoid the interference of initial sheath oscillations on the electron beam energy [50]. This is a standard approach when a voltage is applied between the cathode and anode, see, e.g., Ref. [51,52]. Because electrons emitted from the cathode are accelerated by the cathode sheath, the initial energy of the beam electrons that interact with the bulk plasma $E_{b0}$ equals to $-eV_0$ (small anode sheath voltage of order of $T_{e0}$ can be neglected compared to $-eV_0$). The background pressure was fixed to be the same for all cases $3.85 mTorr$. The pressure is sufficiently low so that the collision frequency between the beam electrons and neutrals, $\nu_{en,elas} \approx 2.1 \times 10^7 s^{-1}$ [53] is small compared to the typical growth rate of the two-stream instability ($\gamma = \sqrt{3}/2\omega_{pe}(n_b/2n_0)^{1/3} \approx 3.02 \times 10^9 s^{-1}$) and therefore, the collisions can be neglected during an initial instability growth period.

Table 1: Transform between dimensional and corresponding dimensionless quantities.

| Quantity | Dimensional Unit | Dimensionless Unit | Normalization Quantity |
|---|---|---|---|
| Time ($t$) | $1 ns$ | $17.82(\omega_{pe} t)$ | $\omega_{pe} \approx 1.78 \times 10^{10} s^{-1}$ |
| Length ($x$) | $1 mm$ | $95.15(x/\lambda_{De,scale})$ | $\lambda_{De,scale} \approx 1.05 \times 10^{-5} m$ |
| Velocity ($v$) | $10^6 m/s$ | $3.76(v/v_{Te,scale})$ | $v_{Te,scale} \approx 2.66 \times 10^5 m/s$ |
| Density ($n$) | $10^{17} m^{-3}$ | $1(n/n_0)$ | $n_0 = 10^{17} m^{-3}$ |
| Electric Field* ($E$) | $10^4 V/m$ | $0.37(E/E_{scale})$ | $E_{scale} \approx 2.69 \times 10^4 V/m$ |
| Energy ($W$) | $1 eV$ | $5(W/T_{e,scale})$ | $T_{e,scale} = 0.2 eV$ |
| Pressure** ($P$) | $1 mTorr$ | $0.26(P/P_{scale})$ | $P_{scale} \approx 3.85 mTorr$ |

*Note: $E_{scale} = \frac{m_e \omega_{pe} v_{Te,scale}}{e}$.

**Note: $P_{scale} = n_g k T_{room}$, $n_g = 1.25 \times 10^{20} m^{-3}$ is the neutral gas density, $T_{room} = 297 K$ is the room temperature.

To classify different regimes of SLT and verify the proposed theory to be described below, we performed 57 simulations with different values of the beam density, beam energy, bulk electron temperature, and gap spacing. The parameters for all the simulation cases are provided in Table A1 in Appendix 3, whereas the parameters of the six simulations studied in detail in the main text are given in Table 2. For each case, the gap spacing is carefully chosen to ensure that the system is sufficiently long to reach wave saturation [54]. All simulations are performed using the previously benchmarked particle-in-cell code EDIPIC-2D, which has been successfully and extensively used to model a variety of plasma devices, see e.g., Ref. [55].

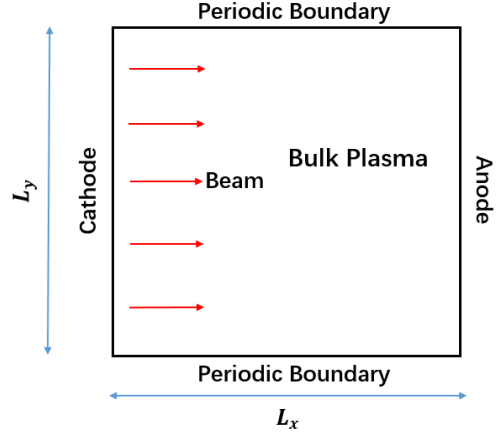

**Figure 1**: Schematic diagram of the simulation domain.

**Table 2**: The six simulation cases studied in detail with parameters: ratio of the beam density over plasma density $n_b/n_{p0}$, the initial energy of beam electrons $E_{b0}$, the initial temperature of bulk electrons $T_{e0}$, and system length in $x$ direction, $L_x$. The parameters for all 57 simulation cases are provided in Appendix 3.

| Case | $n_b/n_{p0}$ | $E_{b0}(eV)$ | $T_{e0}(eV)$ | $L_x(mm)$ |
|---|---|---|---|---|
| 1 | 0.015 | 30 | 0.2 | 32 |
| 2 | 0.00068 | 45 | 0.2 | 32 |
| 3 | 0.015 | 30 | 2.0 | 32 |
| 4 | 0.0015 | 25 | 0.5 | 32 |
| 5 | 0.015 | 80 | 2.0 | 64 |
| 6 | 0.015 | 30 | 0.5 | 32 |

## 3. Simulation Results and Discussions

Nonlinear beam-plasma interaction creates the Langmuir wave packets, which may be subject to different secondary wave-wave instabilities. To better clarify the physical scenarios, we first introduce the key characteristics of four typical regimes of beam-plasma interactions observed in our simulations in the next sub-section 3.1. Case-by-case description for three of the four regimes and their respective dominant instabilities are provided in the following sub-sections 3.2-3.4. Discussions of the electron scattering, turbulence intermittent behavior, the electron energy distribution functions, and energy spectrum are given in sub-section 3.5.

**3.1 Key characteristics of four regimes of beam-plasma interaction**

We characterized and named these four regimes after their respective dominant wave-wave instabilities, which are Parametric Decay Instability (PDI), Standing Wave Modulation Instability (SWMI), Electron Modulation Instability (EMI) and No-MI/PDI regime. The thresholds and growth rates for the instabilities and the key processes in these four regimes are summarized in Table 3. Fig. 2 summarizes the different regimes of beam-plasma interaction observed in our simulations. In Fig. 2, the yellow, blue, and red lines show the thresholds of PDI, SWMI, and EMI, respectively and will be further discussed in Section 4. The whole parameter space is then divided into four regimes:

i) *No-MI/PDI regime:* In this regime that is located below the yellow lines in Fig.2, there are no modulational or parametric decay instabilities. The nonlinear wave coupling is very weak, and the Langmuir wave packets simply saturate through particle-wave trapping mechanism. The nonlinear wave-wave effects in this regime are weak and not important, and this regime has been extensively discussed in previous publications, see e.g., Refs. [3,56,57], therefore no detailed discussion is provided in this paper.

ii) *PDI regime:* This regime corresponds to space between the yellow and blue lines, where the PDI plays a dominant role. The key characteristic for this regime is the generation of the backward waves. Because the PDI growth rate is usually small ($< \omega_{pi}$), the interaction between waves is still weak, resembling the weak turbulence. We therefore also call this regime "Weak Turbulent Regime". Many studies have explored this regime using theory [58-60], experiments [61-63], and numerical simulations [25,26,29,64-67]. However, most of the simulations were limited at applying weak turbulence theory to a weak electron beam-plasma interaction in periodic systems, whereas a more relevant for experiments simulation requires a full kinetic simulation with beam injected from a cathode, which will be given in this paper.

iii) **SWMI regime:** This regime corresponds to parameter space between the blue and the red lines in Fig. 2, where the SWMI becomes dominant. As we will show in Section 4, this instability determines the boundary of "Strong Turbulent Regime". In this regime, standing Langmuir wave packets are generated due to SWMI. The SWMI grows on the ion time scale (the growth rate $\sim\omega_{pi}$). In this process, thermal pressure force created by ion density perturbations is nearly balanced by the ponderomotive force. Despite some experimental [38,68-70] and kinetic simulation studies [71-73] were reported for this regime, most simulations used a wave packet set as an initial condition and the mutual interaction between the beam and wave packet was not modeled. The data resolution for both previous experimental and numerical studies was low due to the limited range of time scales and wavelengths that could be measured at that time [38,68-70]. The accurate threshold of this regime for a beam-plasma system was not identified before, which has caused some confusion in the literature.

iv) **EMI regime:** EMI regime corresponds to parameters space above the red line, where a newly identified modulational instability—EMI becomes dominant. Although the standing Langmuir wave packets are also seen in this regime (similar to the SWMI regime), there are qualitative differences between these two. In the SWMI regime, charge quasi-neutrality condition approximately holds and the growth of the standing Langmuir wave packets is mainly constrained by the balance between the ponderomotive force and the thermal pressure force created by the induced plasma density perturbation. However, in the EMI regime, the growth of the standing Langmuir wave packets is mainly constrained by the balance between the ponderomotive force and the non-ambipolar electron response to the ponderomotive field. EMI grows and saturates on a faster time scale, with the growth rate fast compared with $\omega_{pi}$, correspondingly, the ions dynamics decuple from the wave growth in the initial stage of instability. In contrast to the SWMI which is always accompanied by a large ion density perturbation, the EMI exhibits the rapidly-growing standing Langmuir waves while ions are barely perturbed. It is the electrostatic force resulting from charge separation that balances the strong ponderomotive force (see Section 3.2 and Section 4). The ions respond with a delay on a slower time scale after the field has grown to a significant value. Due to the fast-growing nature of EMI, the resulting electron heating and beam scattering are much stronger in this regime than for the SWMI and PDI regimes. Note that in both SWMI and EMI regimes, large-amplitude wave packets are formed near electron injection location, indicating the occurrence of strong Langmuir turbulence (strong turbulent regime as opposed to a weak turbulent regime).

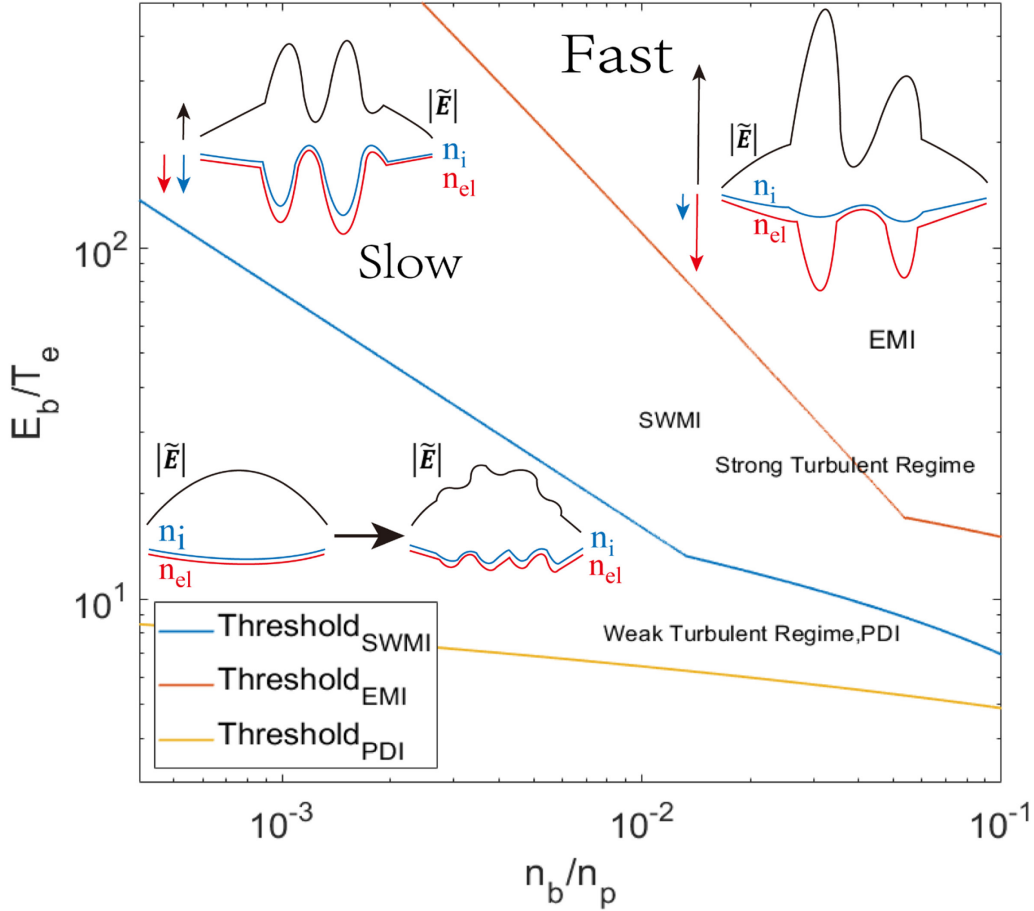

**Figure 2**: Schematic diagram of different regimes in the parameter space of ratio of the beam energy to the bulk electron temperature, $E_b/T_e$, versus the ratio of the beam to plasma density, $n_b/n_p$. The yellow line shows the threshold of Langmuir PDI; the blue line - the threshold of SWMI; the red curve - the threshold of EMI. The analytical relations for the lines are given in the theory part

of Sect. 4 and Appendix 1. In the weak turbulence regime, the wave packet undergoes only slight modulations but doesn't grow locally. In the strong turbulent regime dominated by SWMI, the wave packet grows locally on the ion time scale and ion density perturbations grow together with the wave. Above the threshold of EMI, the wave localization is faster than the ion response and the wave grows locally before ions have time to move. Drawing for the ion density (bottom) and electric field amplitude (top) schematically show these processes for PDI, SWMI and EMI.

Table 3: Summary of the four regimes (EMI, SWMI, PDI, No-MI/PDI) including thresholds, the growth rates of wave-wave interaction, and key characteristics.

| Regime | Threshold[*] | Growth Rate | Key characteristics |
|---|---|---|---|
| **EMI** | Eq. (17) | $> \omega_{pi}$ | Rapidly-growing localized standing waves |
| **SWMI** | Eq. (14) | $\sim \omega_{pi}$ | Localized standing waves |
| **PDI** | Eq. (22) | $< \omega_{pi}$ | Generation of backward waves |
| **No-MI/PDI** | --- | --- | little wave-wave coupling |

[*]See further details in Section 4 and Appendix.

## 3.2 Electron Modulational Instability Regime

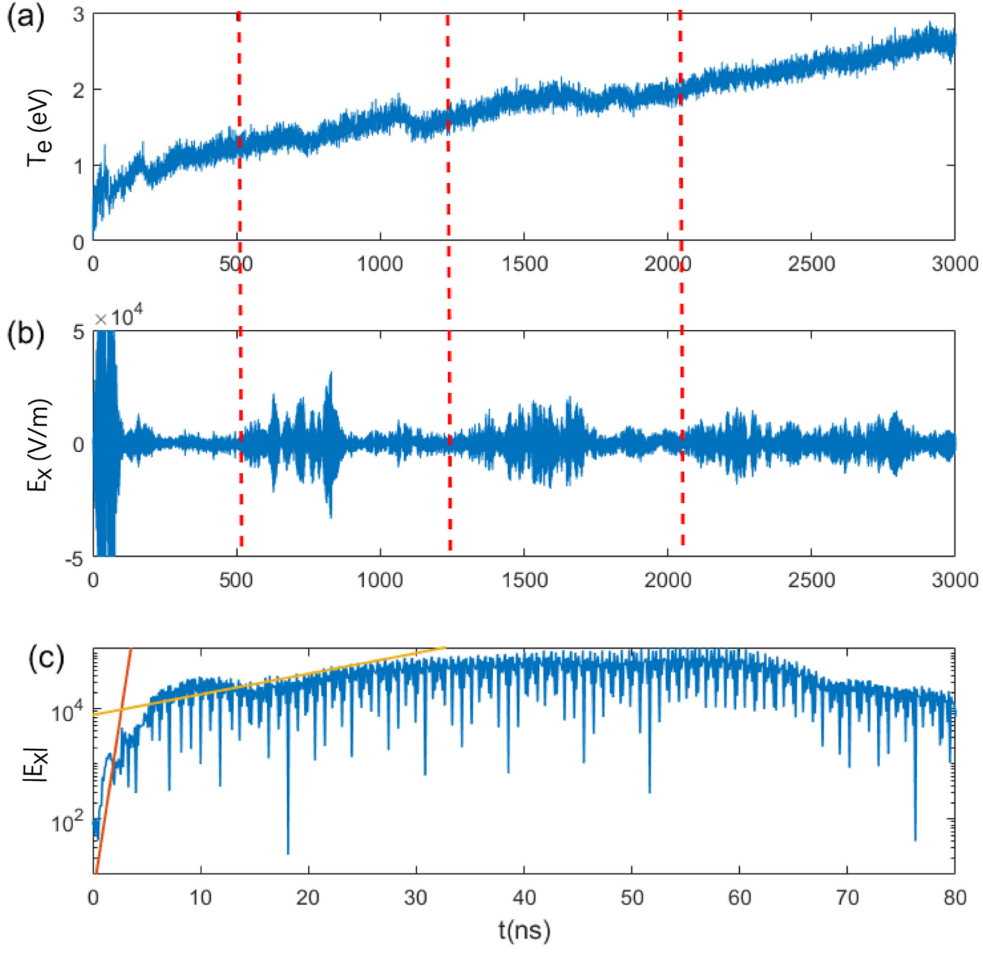

**Figure 3**: Fig.3 (a) shows the time evolution of averaged bulk electron temperature $T_e$. Fig.3 (b) shows the evolution of the electric field, which is output of the EDIPIC probe diagnostics at $x = 2.8mm$, $y = 4.5mm$, where there is intense standing Langmuir wave. The intermittent process can be observed. The three periods are roughly indicated by the red dashed lines. Fig.3 (c) displays the initial evolution of $|E_x|$. The red line shows the linear growth rate of two-stream instability, and the yellow line shows the theoretical growth rate of EMI calculated by Eq. (19) in the Section 4.

First, we illustrate the evolution of waves and particle dynamics for a system initially in the EMI regime based on the simulation results of Case 1 (see Table 2 for the parameters). Fig. 3 shows the temporal evolution of the electron temperature and electric field at the position where standing Langmuir wave packet appears ($x = 2.8mm$, $y = 4.5mm$) in Case 1. We observed an intermittent process with quasi-periodic instability bursts [marked by the red dashed lines in Fig.3(a), 3(b)], repeating itself with a nearly constant period ($T < 750 ns$) exhibiting the intermittent nature of turbulence in this regime [74].

From Figs. 3 (a) and 3 (b), it is seen that the electron temperature gradually increases while the amplitude of the burst electric field decreases. The increase in electron temperature indicates the electron heating by the waves. Such an increase in temperature in turn diminishes the intensity of the wave-wave interactions, and brings the system from EMI to SWMI regime since the second burst. Eventually, the periodic burst feature stops because the bulk electron temperature becomes sufficiently high. Note that the heating efficiency and intermittent feature also depends on system size and beam width, a longer system with narrower beam could stay in EMI regime for a long time (see Section 3.5.2). Fig. 3(c) shows the evolution of the amplitude of the electric field at the initial instability-growing stage. As we can see, the linear stage of beam-plasma instability lasts only for approximately $4 ns$. After that, the growth rate becomes smaller. At around $t = 20 ns$, a secondary instability comes into play and the amplitude further increases until $t = 60 ns$. This instability is referred to as the EMI. The theoretical growth rate of EMI is $\gamma_{EMI} \approx 7.4 \times 10^7 s^{-1} > \omega_{pi} \approx 3 \times 10^7 s^{-1}$, which is faster than the ion response and agrees well with our theory [Eq. (19)].

Figure 4 shows the evolution of the plasma parameters in Case 1 at different moments. Here, we only show the first burst period to illustrate the evolution of the strong Langmuir turbulence. Fig 4 (a) shows the time evolution of the averaged electric field energy $\epsilon_{E,mean}$, the averaged kinetic energy for bulk electrons $\epsilon_{K,mean}$, the averaged energy transfer rate from wave to beam, $E \cdot J_{beam}$, and to the bulk plasma, $E \cdot J_{bulk}$. As indicated by the dashed lines, each burst period can be roughly divided into three stages. The beam electrons are accelerated by the cathode sheath [see the arrows in Fig. 4 (b2)] and then interact with the background plasma, creating a large-amplitude Langmuir wave packet [see Fig. 4 (b1)]. In Stage I, the electric field energy increases dramatically during $t = 0 - 60 ns$ and decreases during $t = 60 - 90 ns$. At $t =$ $60ns$, the amplitude of the Langmuir wave packet has grown to $E^2 \sim 6 \times 10^9 V^2/m^2$, which is much higher than the typical saturation level of the two-stream instability ($E^2 \sim 9/4 n_0 m_e v_b^2 (n_b/n_0)^{4/3}/\epsilon_0 \sim 1.08 \times 10^9 V^2/m^2$).

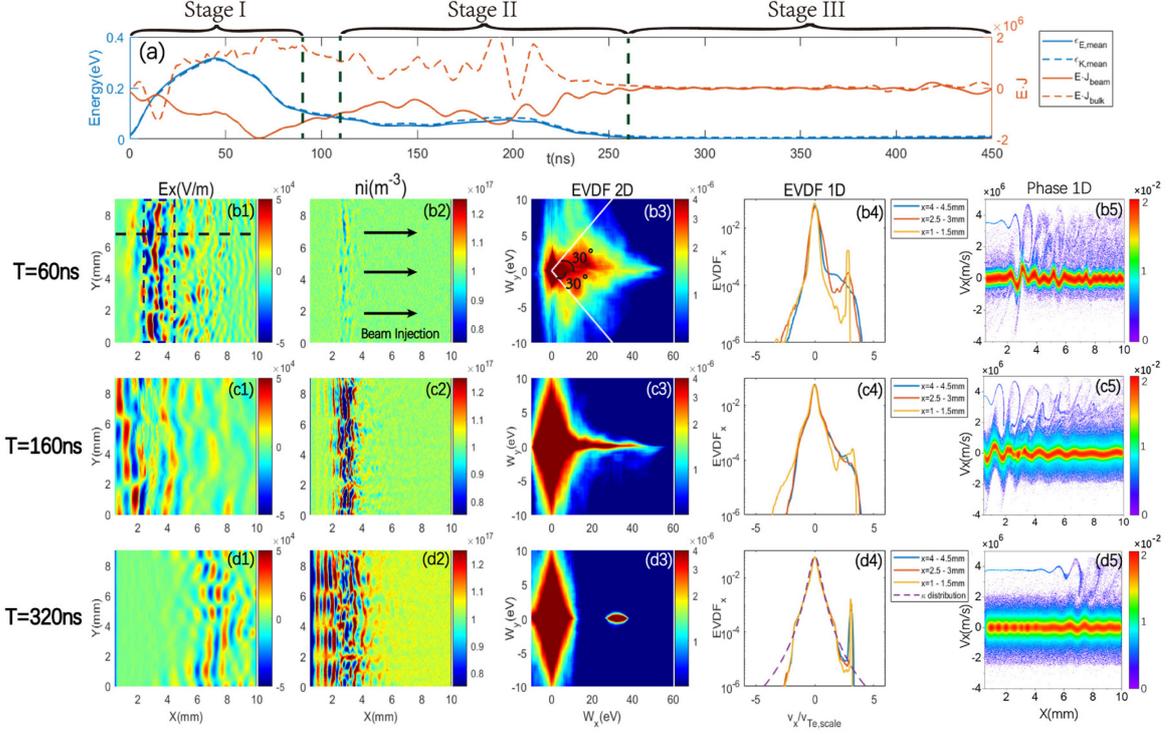

**Figure 4**: Evolution of turbulence and electron heating for Case 1. Blue lines in Fig.4(a) show the evolution of the averaged electric field energy, $\epsilon_{E,mean}$ (solid line) and the averaged kinetic energy for bulk electrons, $\epsilon_{K,mean}$, (dashed line), during $t = 0 - 450 ns$. Red lines in Fig.4 (a) show the averaged energy transfer rate from wave to beam, $E \cdot J_{beam}$, (solid line) and to the plasma, $E \cdot J_{bulk}$, (dashed line). $E \cdot J_{beam}$ has a negative value indicating that the beam energy is transferred to waves; $E \cdot J_{bulk}$ always has a positive value, indicating that waves are transferring energy to the plasma. The averages are taken over $x = 2.5 - 4.5 mm$ (shown by the blue dashed rectangle in Fig.4 (b1)). Figs 4(b)-(d) show color plots of $E_x$, $n_i$ and EVDF at different times. The 2D EVDFs for $f(w_x, w_y)$ are averaged over region $x = 4 - 5mm$, $y = 1 - 8mm$. The 1D EVDFs $f(v_x)$ are averaged over region $y = 1 - 8mm$ and $x$ regions denoted in the figure legend. <span style="color:red">Each EVDF is normalized to unity by the integration of EVDF.</span> The purple dashed line in Fig.4 (d4) shows the kappa distribution with coefficients $\kappa = 1.6$ and $T = 1.0 eV$. The data for the 1D phase space plot is taken along a fixed $y$ value, the black dashed line $y = 6.8mm$ in (b1).

Two important features of the EMI regime are observed in simulations. First, the wave energy grows and saturates, during which a solitary-like electric field waveform forms, indicating a "localization" of Langmuir waves faster than ion response [see Fig. 4(b1)]. Second, the electric field is strongly localized while the ion density is almost uniform. *This is in contrast to the traditional Langmuir collapse scenario* [70,75,76] where the ion density is strongly perturbed and the thermal pressure force $\nabla(\delta n_e T_e)$ was assumed to nearly balance the ponderomotive force $\nabla(\epsilon_0 E^2/4)$. Instead, it is the electrostatic force $n_{p0} e E_l$ (where $\nabla E_l = e/\epsilon_0 (n_i - n_{el})$, "$l$" denotes time average) resulting from charge separation that balances the ponderomotive force $\nabla(\epsilon_0 E^2/4)$ (see Section 4 and our short Letter [49]). The movies showing the evolution of Langmuir wave energy as well as the associated ion density perturbations are available in the supplementary material [77] and better show that the growth of the wave packet is faster than the ion response, indicating that this is not a classic Langmuir collapse process. These unique features of EMI could not be modelled by traditional Zakharov equations [33,78] and was not studied in detail by previous beam-plasma experimental studies [38,68,70,79]. From Fig. 4(a) it is evident that $E \cdot J_{bulk}$ increases when the electric field begins to decrease, indicating a strong energy transfer from the wave electric field to bulk electrons, which is known as "burnout" of the wave packet [1,72]. As a result, the average temperature in Case 1 increases from 0.2eV to 1.07eV during $t = 0 - 90ns$ (see Fig. 3). The super-thermal electrons accelerated by the intense electric field can also be seen in 1D EVDF plot in Fig. 4 (b4). At the same time, the beam is scattered by the plasma waves off the central line ($W_y = 0eV$) by the strong electric fields parallel to the electrode, in the energy range from $W_y \sim -5eV$ to $W_y \sim 5eV$. The scattering angle could be as large as $\theta = \arctan v_y/v_x = 30°$ (see the white lines in Fig. 4 (b3)). The backward electron jet formation in the phase space plot given by Fig. 4 (b5) also indicates a large amplitude standing wave. Because ions are heavy, it takes time for ions to respond to the ponderomotive force. At about $t = 110ns$, the initial ion density perturbation grows to a significant value $\delta n_i/n_{i0} \sim 0.49$, marking the beginning of Stage II [Fig. 4(c2)]. The fact that $\delta n_{i,max}/n_{i0} \sim 0.59 < \epsilon_0 |E_{peak}|^2/4n_{p0}T_e \sim 2.5$ further confirms that the ions do not have enough time to respond to the fast growing wave so that the thermal pressure cannot balance the ponderomotive force (meaning that it is not a traditional Langmuir collapse process). In Stage II, the amplitude of the Langmuir waves dramatically decreased. As the initial ion density perturbation is large, it acts like a virtual wall, which tends to reflect the Langmuir waves and create the new secondary standing wave and the associated secondary density perturbation at around $x < 2mm$ [see Fig. 4(c1) and 4(c5)]. Also, in this stage, the initial ion density perturbation spreads to the surrounding regions in the form of the ion acoustic waves [73]. At the same time, the electron beam relaxes quickly and becomes much less scattered [see Figs. 4(c3) and 4(c4)]. When the secondary ion density perturbation grows to about 30% near the injection point at $x < 2mm$, the Stage III starts ($t > 260ns$).

In Stage III, a region of strong ion density perturbations near the beam injection point (see Fig. 4(d2)) modifies the beam-plasma interaction: $E_y$ becomes almost zero in the region with strong ion density perturbations in region $x = 0 - 4mm$; and the beam does not interact with waves until it passes through the region with strong ion density perturbation (Fig. 4(d1)). The decoupling of beam from the plasma can also be seen from the evolution of the EVDFs shown in Figs. 4(d3) and 4(d4), the evolution of transferred energy shown in Fig. 4(a) as well as the phase space plot Fig. 4(d5). Interestingly, we observed the formation of a well-fitted *kappa distribution* in Fig. 4(d4), with a coefficient $\kappa = 1.6$ and $T = 1.0eV$. This means that the strong turbulence efficiently produces energetic electrons since it implies that the electrons with lower velocities are also rapidly interacting with waves with high wave numbers ($v_e \sim \omega_{pe}/k$, see also Appendix 4).

Figure 5 shows the wave energy spectra for Case 1. The generation of a standing wave can be seen in Fig. 5(a), which is a clear indication of strongly nonlinear wave-wave interactions. A very broad spectrum with the presence of strong backscattered plasma waves (negative $k$) can be observed in Fig. 5(b). The vertical red dashed line marks the wave-vector $k_{max,1} \approx 9.1 \times 10^3 m^{-1}$, which corresponds to the largest wave energy. For Case 1, we have $k_{max,1} > k_0 = \omega_{pe}/v_b \approx 5.5 \times 10^3 m^{-1}$, where $k_0$ is the pump scale, meaning that the wave energy tends to spread to smaller scales (larger $k$). The formation of such a broad spectrum can be understood from a theoretical analysis performed by applying the Fourier transform to the wave packet equation (see Appendix 1), and describes a four-wave process between the original Langmuir wave and daughter waves $(L_{up}, L_{low})$ ($2L_0 \rightarrow L_{up} + L_{low}$). In linear approximation it satisfies $2k_0 = k_{up} + k_{low}$, therefore this process tends to populate the spectrum to higher $k$ since $L_{up}$ contains more energy than $L_{low}$. As the modulation instability grows and the wave energy is localized, near beam injection standing waves generate a strong plasma density perturbation via the action of the ponderomotive force in the later nonlinear stage (after $t = 45ns$ for Case 1); the lower-frequency density perturbation interacts with the initial pump wave, generating stronger backscattered wave and amplifing standing wave, which in turn increases the ponderomotive force and yields further growth of the density perturbation. This self-sustained nonlinear process can explain the results for Case 1. Besides, we observe a power law spectrum $E^2 \sim k^{-5}$ for $k > k_0 \approx 5.5 \times 10^3 m^{-1}$ (see Fig. 5(d)). Such a spectrum can also be seen in other simulation cases during the nonlinear Langmuir wave growth caused by the Standing Wave Modulational Instability. The exact reason for the formation of such a spectrum remains unclear. We provide one possible explanation in Appendix 4.

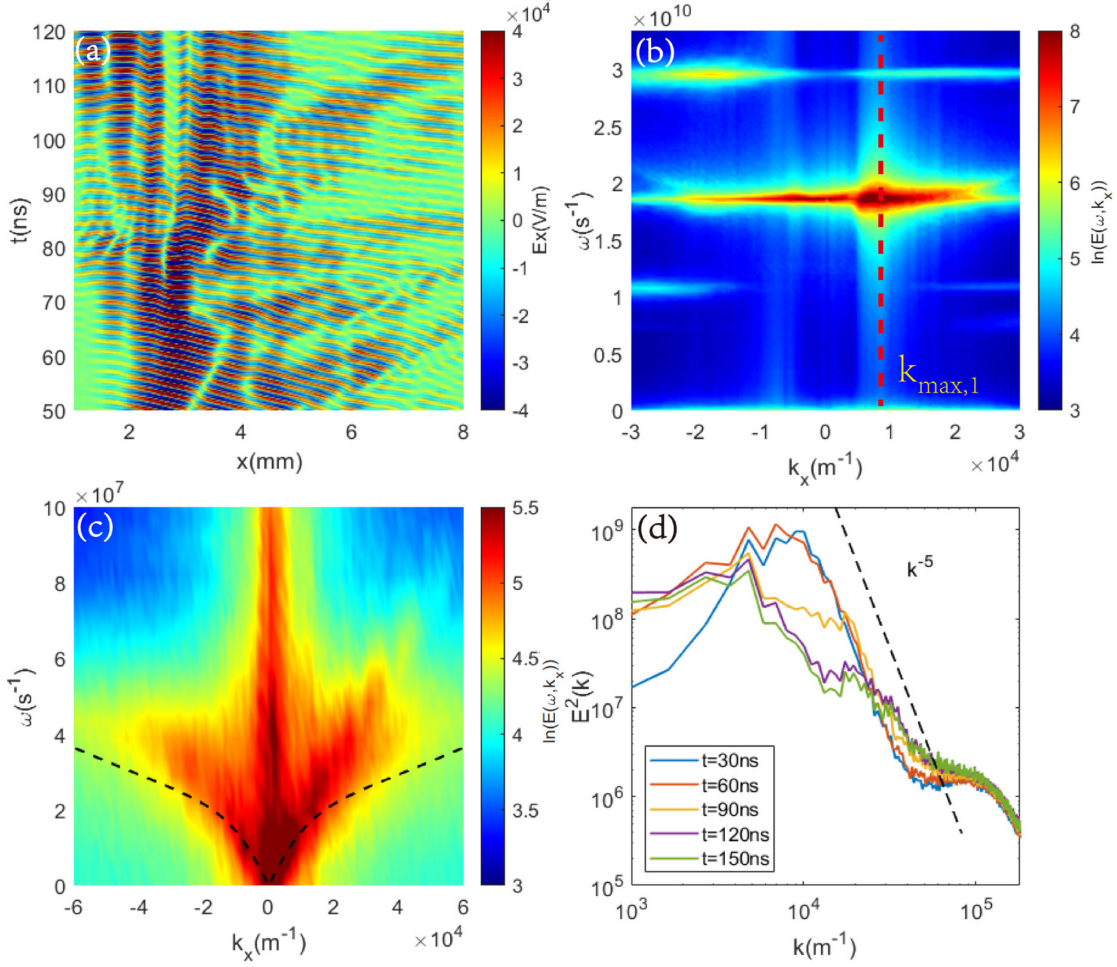

**Figure 5**: Properties of strong turbulence produced by the beam for Case 1. Fig.5(a) shows the 2D color plot of the parallel electric field $E_x$ along $y = 6.8mm$ in Fig. 4 as a function of $x$ and $t$ during $t = 50 - 120ns$. Fig.5 (b) shows the high-frequency spectrum of plasma waves corresponding to the dispersion relation of Langmuir ($L$) waves. The dispersion relations are obtained by performing 3D FFT during $t = 50 - 80ns$ and then summing over $k_y$ space. The $k_{max,1} \approx 9.1 \times 10^3 m^{-1}$ therein denotes the k wave-vector with the most wave energy. Fig.5 (c) shows the spectrum of low-frequency waves corresponding to the dispersion relation of ion acoustic waves ($IAW$), which is also obtained from 3D FFT, but during $t = 0 - 450ns$. The black dashed lines therein indicate the dispersion relation of ion acoustic waves. Fig.5 (d) shows the 1D energy spectrum $E^2(k)$ at different times, where $k = \sqrt{k_x^2 + k_y^2}$ and a clear $-5$ power spectrum can be seen at $t = 30ns, 60ns$.

## 3.3 Standing Wave Modulational Instability Regime

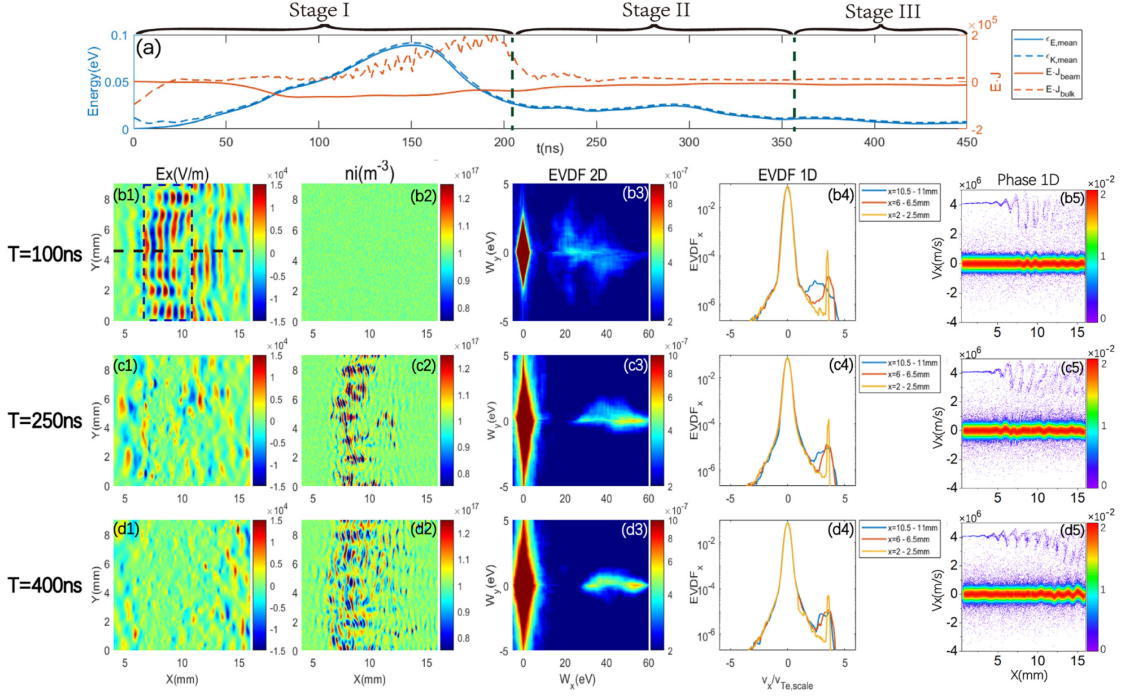

**Figure 6**: Evolution of turbulence and electron heating for Case 2, similar to Fig.4. Blue lines in Fig.6(a) show the time evolution of the averaged electric field energy, $\epsilon_{E,mean}$ (solid line), and the averaged kinetic energy for bulk electrons, $\epsilon_{K,mean}$ (dashed line), during $t = 0 - 450ns$. Red lines in Fig.6 (a) show the averaged energy transfer rate from wave to beam, $E \cdot J_{beam}$ (solid line) and to the plasma bulk component, $E \cdot J_{bulk}$ (dashed line). $E \cdot J_{beam}$ has a negative value indicating that the beam energy transferring to waves; $E \cdot J_{bulk}$ always has a positive value, indicating that waves are transferring energy to the bulk component. The averages are taken over $x = 7 - 11mm$ (shown by the blue dashed rectangle in Fig.6 (b1)). (b)-(d) show a snapshot of the system at different times. The 2D EVDFs for $f(w_x, w_y)$ are averaged over region $x = 10.5 - 11.5mm$, $x = 1 - 8mm$. The 1D EVDFs $f(v_x)$ are averaged over region $y = 1 - 8mm$ and $x$ regions denoted in the figure legend. Each EVDF is normalized to unity by the integration of EVDF. The data for the 1D phase space plot is taken along a fixed $y$ value, the black dashed line $y = 4.5mm$ in (b1).

Fig. 6 shows the time evolution of turbulence for Case 2 in Table 2. As we can see in Fig. 6(a), the evolution of electric field energy exhibits three stages, similar to Case 1. However, one major difference is that, the growth rate of field amplitude $\gamma \approx 1.43 \times 10^7 s^{-1} \sim \gamma_{SWMI} = 2.2 \times 10^7 s^{-1} \sim \omega_{pi}$ (where the $\gamma_{SWMI}$ is the theoretical growth rate of SWMI given by Eq. (15) in Section 4), which means that the instability grows on the ion time scale, while for Case 1 EMI instability developed faster than the ion response frequency. The time when field energy reaches maximum for Case 2 (~$150ns$) is much longer than that in Case 1 (~$50ns$), comparing Fig. 6(a) with Fig. 4(a). In contrast to the EMI regime, in the initial stage of the SWMI, the thermal pressure nearly balances the ponderomotive force $\nabla(\delta n_e T_e) \approx \nabla(\epsilon_0 |E|^2/4)$ and is confirmed by simulation data (see Section 4). In Stage I shown by Figs. 6(b1)-6(b4), a quasi-stationary wave envelope is formed via electron-beam plasma interaction at around $t = 80ns$. Since the electric field amplitude is greater than the threshold of SWMI (see Section 4), SWMI development leads to formation of localized peaks, while the ponderomotive force action tends to produce ion density holes. In Stage II, strong ion density perturbations have formed. Multiple ion density depletions can be seen in Fig. 6(c2), indicating the formation of multiple standing waves. *This phenomenon is different from previous experimental observations* [38]*, where multiple peaks form only when the beam density is sufficiently large ($n_b/n_p > 0.6\%$)*. The difference may be due to the small transverse size of the electron beam used in that experiments (as opposed to uniform in transverse direction beam used in our simulations). No clear secondary standing waves and secondary ion density perturbations are observed in the Case 2 in contrast to Case 1. In Stage III, at $t = 400ns$, the initial ion density perturbations have spread to a larger region, causing the decrease of transfer energy from beam to the waves (see Fig.6(d1) and Fig.6 (d2)). The decoupling can be also noticed from the EVDF plots in Figs. 6(b3)-6(d3), Figs. 6(b4)-6(d4) as well as the phase space plot Figs. 6(b5)-6(d5). The bulk electron temperature increases from $0.2eV$ to $0.41eV$, which is small compared with that in Case I (where the temperature increases from 0.2eV to 1.07eV), indicating that SWMI is less efficient in electron heating than EMI.

## 3.4 Parametric Decay Instability Regime

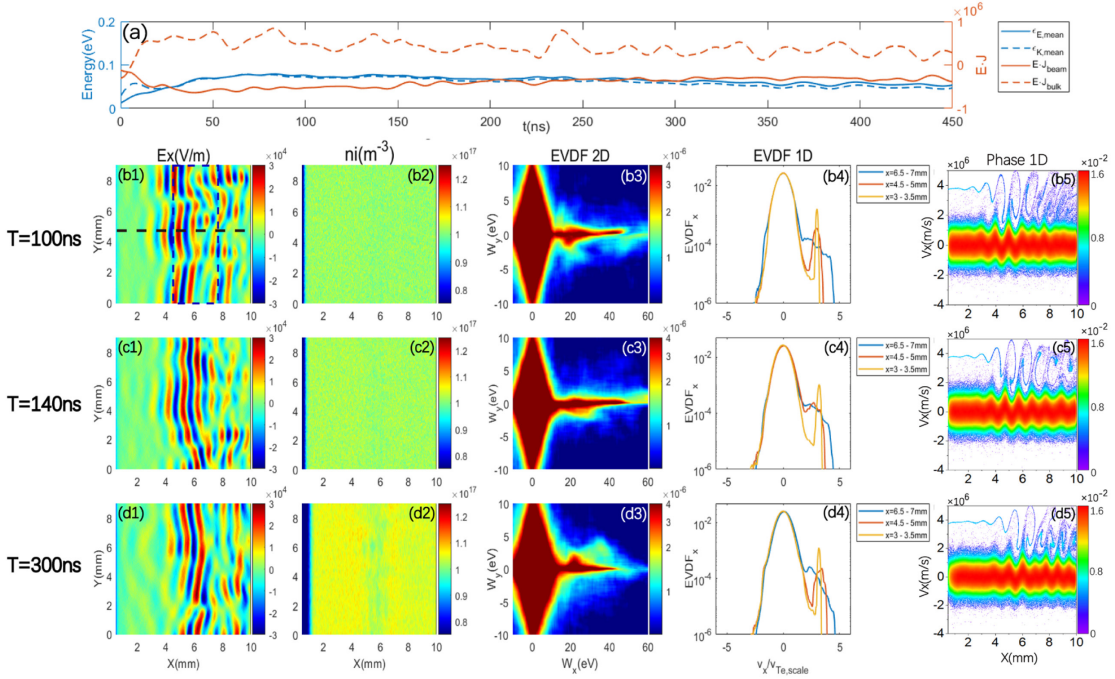

**Figure 7**: Evolution of turbulence and electron heating for Case 3, similar to Fig.4. Blue lines in Fig.7(a) show the time evolution of the averaged electric field energy, $\epsilon_{E,mean}$ (solid line), and the averaged kinetic energy for bulk electrons, $\epsilon_{K,mean}$ (dashed line), during $t = 0 - 450ns$. Red lines in Fig.7 (a) show the averaged energy transfer rate from wave to beam, $E \cdot J_{beam}$ (solid line) and to the plasma bulk component, $E \cdot J_{bulk}$ (dashed line). $E \cdot J_{beam}$ has a negative value indicating that the beam energy transferring to waves; $E \cdot J_{bulk}$ always has a positive value, indicating that waves are transferring energy to the bulk component. The averages are taken over $x = 4.5 - 7.5mm$ (shown by the blue dashed rectangle in Fig.7 (b1)). (b)-(d) show a snapshot of the system at different times. The 2D EVDFs for $f(w_x, w_y)$ are averaged over region $x = 6.5 - 7.5mm$, $x = 1 - 8mm$. The 1D EVDFs $f(v_x)$ are averaged over region $y = 1 - 8mm$ and $x$ regions denoted in the figure legend. Each EVDF is normalized to unity by the integration of EVDF.

The data for the 1D phase space plot is taken along a fixed $y$ value, the black dashed line $y = 4.5mm$ in (b1).

Figure 7 shows the time evolution of the plasma parameters in Case 3 in Table 2 (weak turbulence case). Please note that Case 3 in this paper is Case 2 in our short Letter [49]. As we can see, the electric field energy is almost a constant value of $0.08eV$ after the saturation of initial growth, while we can see wave energy grows and decays both Case 1 and Case 2. The energy transfer rate is also much smaller than that in Case 1 and Case 2. A wave envelope structure is observed in the electric field profile, but it is nearly stationary. This is in clear contrast to the previous two cases where the wave envelopes are strongly perturbed by the localized Langmuir wave field. The ion density is not modulated until parametric decay instability (PDI) grows to a noticeable value at t $\approx$ 250ns. The growth rate for PDI occurrence $\gamma \approx 4 \times 10^6 s^{-1}$ matches reasonably well with the theoretical growth rate of PDI $\gamma_{PDI} \approx 7.6 \times 10^6 s^{-1}$. We can see from Fig.7(d2) that the ion density is only slightly perturbed and the clear ion density dips in Case 1 and Case 2 are not seen. We also can see that the EVDF and phase space change very little in Figs. 7(b4)-7(d4) and Figs. 7(b5)-7(d5).

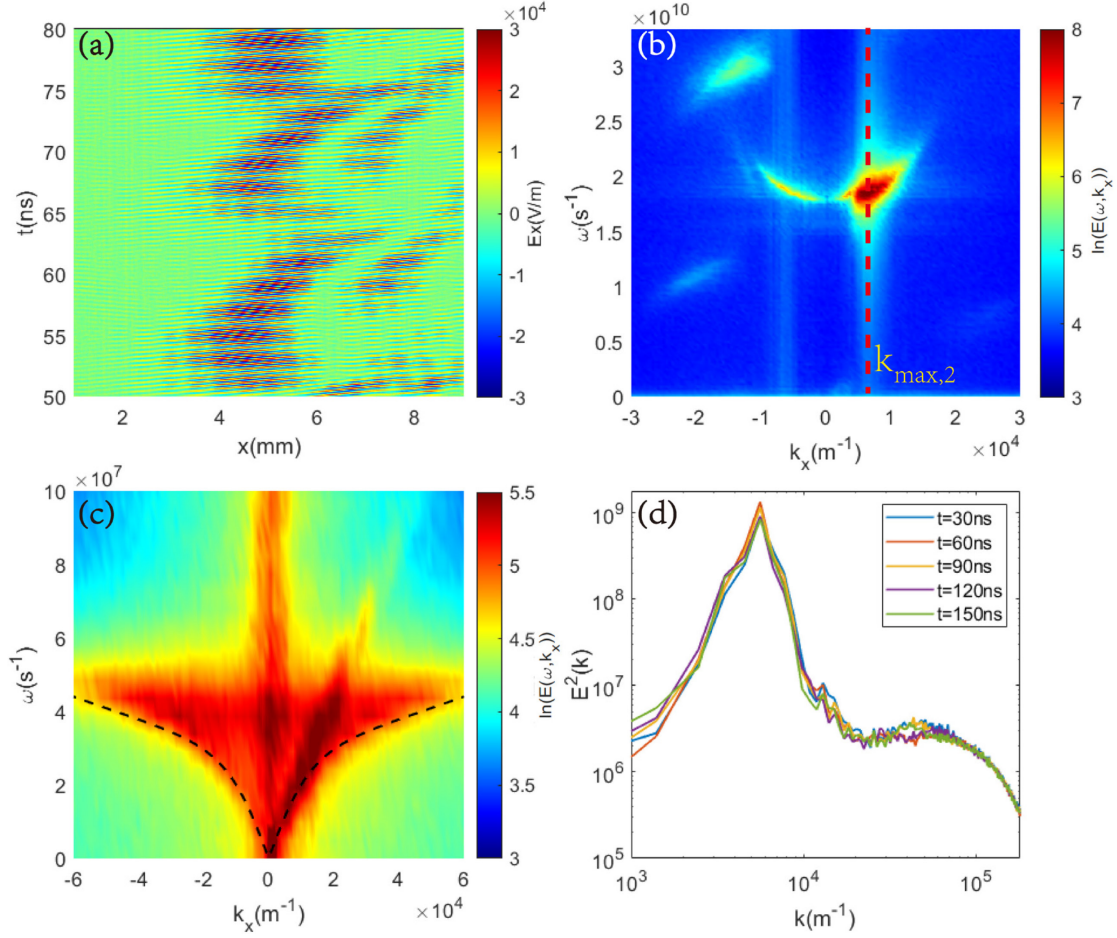

**Figure 8**: Properties of strong turbulence produced by the beam for Case 3. Fig.8(a) shows the 2D color plot of the parallel electric field $E_x$ along $y = 6.8mm$ in Fig. 7 as a function of $x$ and $t$ during $t = 50 - 80ns$. Fig.8 (b) shows the high-frequency spectrum of plasma waves corresponding to the dispersion relation of Langmuir (L) waves. The dispersion relations are obtained by performing 3D FFT during $t = 50 - 80$ns and then summing over $k_y$ space. The $k_{max,2} \approx 5.5 \times 10^3 m^{-1}$ therein denotes the $k$ wave-vector with the most wave energy. Fig.8 (c) shows the spectrum of low-frequency waves corresponding to the dispersion relation of ion acoustic waves (IAW), which is also obtained from 3D FFT, but during $t = 0 - 450$ns. Fig.8 (d) shows 1D energy spectrum $E^2(k)$ at different times, where $k = \sqrt{k_x^2 + k_y^2}$.

Fig. 8 further shows that the forward propagating Langmuir waves dominate [see Figs. 8(a) and 8(b)], with the wave number $k_{max,2} \approx k_0 \approx 5.5 \times 10^3 m^{-1}$. The wave spectrum in the low-frequency range is shown in Fig. 8(c), from which the ion acoustic waves can be observed, indicating the occurrence of PDI for Langmuir waves ($L \rightarrow$ $L' + IAW$, where $L$ and $L'$ represent the forward and backward propagating Langmuir waves, and IAW represents the forward propagating ion acoustic wave [63]) in Case 3 because the three-wave resonance condition can be well satisfied ($\omega_L = \omega_{L'} + \omega_{IAW}$, $\vec{k}_L = \vec{k}_{L'} + \vec{k}_{IAW}$). Fig. 8(d) further confirms that the Langmuir wave energy is almost concentrated at the original pump wave-vector.

Here, Case 4 with lower electron temperature (0.5eV vs 2eV) and lower beam density (0.0015 vs 0.015) in the PDI regime is briefly discussed as well. The purpose of this case is to show the correctness of our threshold given by Eq. (14); the wave behavior in Case 4 is also slightly different from Case 3. Fig. 9 shows the time evolution of Case 4. As we can see in Fig.9(a), the electric field energy does not change greatly after it saturates at 0.04eV until $t = 200ns$ when the PDI grows. The growth rate for PDI occurrence $\gamma \approx 6.7 \times 10^6 s^{-1}$ roughly agrees with the theoretical growth rate of PDI $\gamma_{PDI} \approx 1.06 \times 10^7 s^{-1}$ in this case. The PDI creates a backward propagating Langmuir wave packet, which moves towards the beam injection point, making the electric field in the original location weaker (comparing (b1)-(d1)). That explains the drop in electric field energy in Fig.9(a) (this is because the energy is calculated by taking an average over a fixed region). Because of a larger wavelength of Langmuir waves, in this case, the backward Langmuir waves undergo less damping and propagate for a longer distance compared with Case 3. Despite the weak ion density perturbation observed in Fig. 9(d2), a localized standing wave feature is not observed in this case. From the Fig. 14 below, we can see both Case 3 and Case 4 are very close to the boundary of strong turbulent regime. The physical reason why there is no local wave growing, in this case, is because the wave is not strong enough to overcome the wave dispersion (corresponding to the first component of the threshold of SWMI in Eq. (14)), while in Case 3 the reason is mainly due to strong damping (corresponding to the second component of the threshold of SWMI in Eq. (14)). This further confirms the correctness of the threshold that we obtained from theory part, since both components of the threshold are necessary. The EVDF manifests little changes between different times.

The phase space plot almost doesn't change until the occurrence of PDI (see Figs. 9(b5)-9(d5)). The bulk electrons are heated by the electric field and because of the motion of the wave packet, we see a slight change in the location of beam relaxation (see Fig. 9 (c4)).

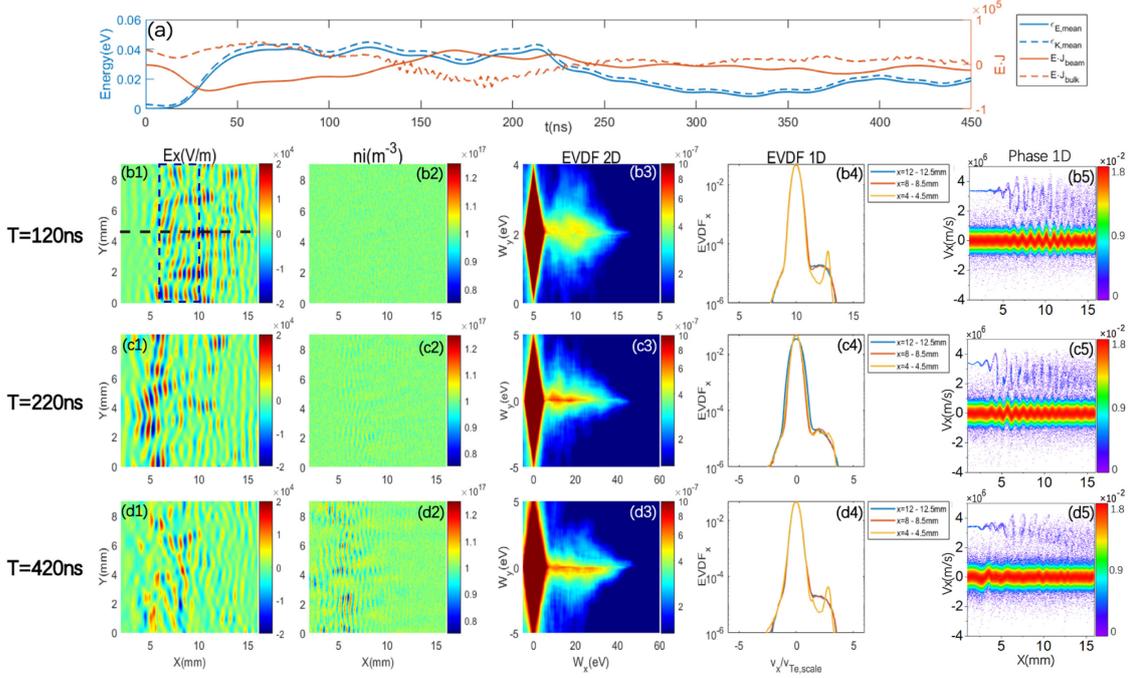

**Figure 9**: Evolution of turbulence and electron heating for Case 4, similar to Fig.4. Blue lines in Fig. 9(a) show the time evolution of the averaged electric field energy, $\epsilon_{E,mean}$ (solid line), and the averaged kinetic energy for bulk electrons, $\epsilon_{K,mean}$ (dashed line), during $t = 0 - 450 ns$. Red lines in Fig. 9(a) show the averaged energy transfer rate from wave to beam, $E \cdot J_{beam}$ (solid line) and to the plasma bulk component, $E \cdot J_{bulk}$ (dashed line). $E \cdot J_{beam}$ has a negative value indicating that the beam energy transferring to waves; $E \cdot J_{bulk}$ always has a positive value, indicating that waves are transferring energy to the bulk component. The averages are taken over $x = 6 - 10 mm$ (shown by the blue dashed rectangle in Fig. 9(b1)). (b)-(d) show a snapshot of the system at different times. The 2D EVDFs for $f(w_x, w_y)$ are averaged over region $x = 12 - 13mm$, $x = 1 - 8mm$. The 1D EVDFs $f(v_x)$ are averaged over region $y = 1 - 8mm$ and $x$ regions denoted in the figure legend. Each EVDF is normalized to unity by the integration of EVDF. The data for the 1D phase space plot is taken along a fixed $y$ value, the black dashed line $y = 4.5mm$ in (b1).

## 3.5 Discussions of simulation results

### *3.5.1 Electron scattering*

Strong Langmuir Turbulence (SLT) also gives rise to a strong electron scattering because of the formation of large perpendicular electric field. A detailed analysis of this complicated process is beyond the scope of this paper. Here we only provide a rough estimate for beam scattering in strong turbulence. Fig. 10 shows the averaged scattering of the beam electrons for different cases. The averaged scattering is defined by: $V_{y,mean}(x) = \sum_{i=1}^{N} |v_{y,i}(x)|/N$, and $V_{y,0} = V_{y,mean}(x_0)$. Here we also used Case 5 in the SWMI regime and Case 6 with EMI to show that there are also strong beam scatterings. Locations $x_0$ are chosen differently for different cases in order to better describe the scattering. As we can see in this figure, the cases with SLT all show a clear beam scattering (see Figs. 10 (a), (c), and (d)), although such a scattering could be the strongest at different times. This is because the wave amplitude reaches its peak value at different moments for different cases. We can also see that the strong scattering could last for a longer time in Case 5 for SWMI than in Case 1 and Case 6 for EMI. The reason is because the strong field in Case 1 and Case 6 for EMI grow faster than SWMI case and could not last for a long time. Cases without strong turbulence show little beam scattering (see Fig. 10 (b) for Case 3), even if PDI is occurring in the system. The physical reason for this result could be understood by the fact that a simple plane wave with the form $E_0 e^{i(kx-\omega t)}$ does not produce scattering of the beam, while in the nonlinear strong turbulence the Langmuir waves form a strongly localized standing wave structure, whose ponderomotive force could expel electrons out of that localized region, creating net scattering to the beam particles around it.

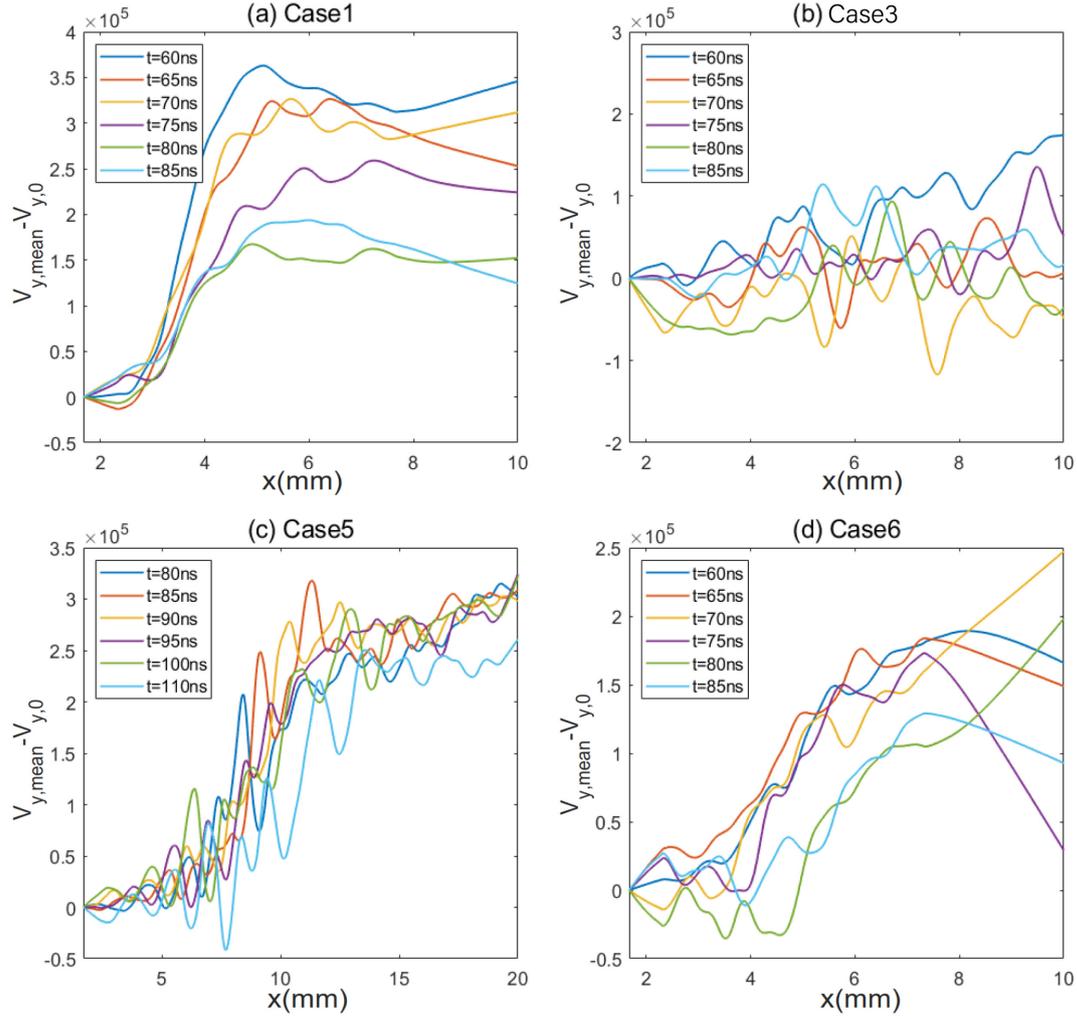

**Figure 10**: Averaged $|v_y(x)|$ of the beam as a function of $x$ showing beam scattering at different times for (a) Case 1, (b) Case 3, (c) Case 5, and (d) Case 6.

### *3.5.2 Intermittent feature*

We showed via 2D-PIC simulations that the SLT produced by electron beam-plasma interaction with a continuous beam injection is intermittent with periodic bursts. It is worth mentioning that intermittent feature of strong Langmuir turbulence has been observed in previous fluid simulations using the Zakharov equations in periodic systems [78] as well as in experimental studies [79]. The repetition period that we observed here in Case 1 ($< 750$ns, corresponding to $\omega_{pe}T < 10^4$) is short, indicating that a system initially in EMI regime in our case also undergoes rapid, intense burst periods.

From Case 1, we see that as the electron temperature grows due to heating, the system moves from EMI regime to SWMI regime and finally to PDI regime. The Langmuir wave packet gradually becomes stable to both SWMI and EMI, and the periodic burst feature disappears. *However, we emphasize that such an intermittent feature, as well as the time duration that a system could stay in the EMI regime, depends strongly on various other factors, including the system length and beam width.* A longer system contains more electrons, making electron heating less efficient. Similarly, for a narrow beam, the transverse energy spread can reduce electron heating by beam. We performed simulations with a narrower beam as compared with Case 1. The results are shown in Fig. 11. We can see from Fig. 11 (a) that the temperature increased less rapidly and intermittent feature became more significant (the amplitude decrease between two consecutive bursts becomes smaller, the periodic burst feature could last for a longer time). It is worth mentioning that there is still significant charge non-neutrality even in the second burst. Therefore, the validity of our simulations and analytical theory may not be just limited to the initial stage of a system. They could also be implemented to understand a pulsed beam system such as in the hollow cathode [48] and in the electron beam-generated plasma [6,45-47] where the plasmas are usually cold ($T_e \leq 1eV$ because the vibrational and rotational excitation energy of nitrogen gas cools the plasma electrons and strongly reducing $T_e$) and the beam could be very intense (in Ref. [45-47] the authors used pulsed keV electron beam). This observation might also explain the burst Langmuir wave packets observed in the space plasmas, such as Earth's Auroral ionosphere [12], and Jupiter's bow shock [14] for infinitely long plasmas. For the case of type III solar radio bursts [80], the electron beam is believe to be generated inside the plasma, an electromagnetic periodic system with electron beam should be a more appropriate modelling, which will be left for future investigations.

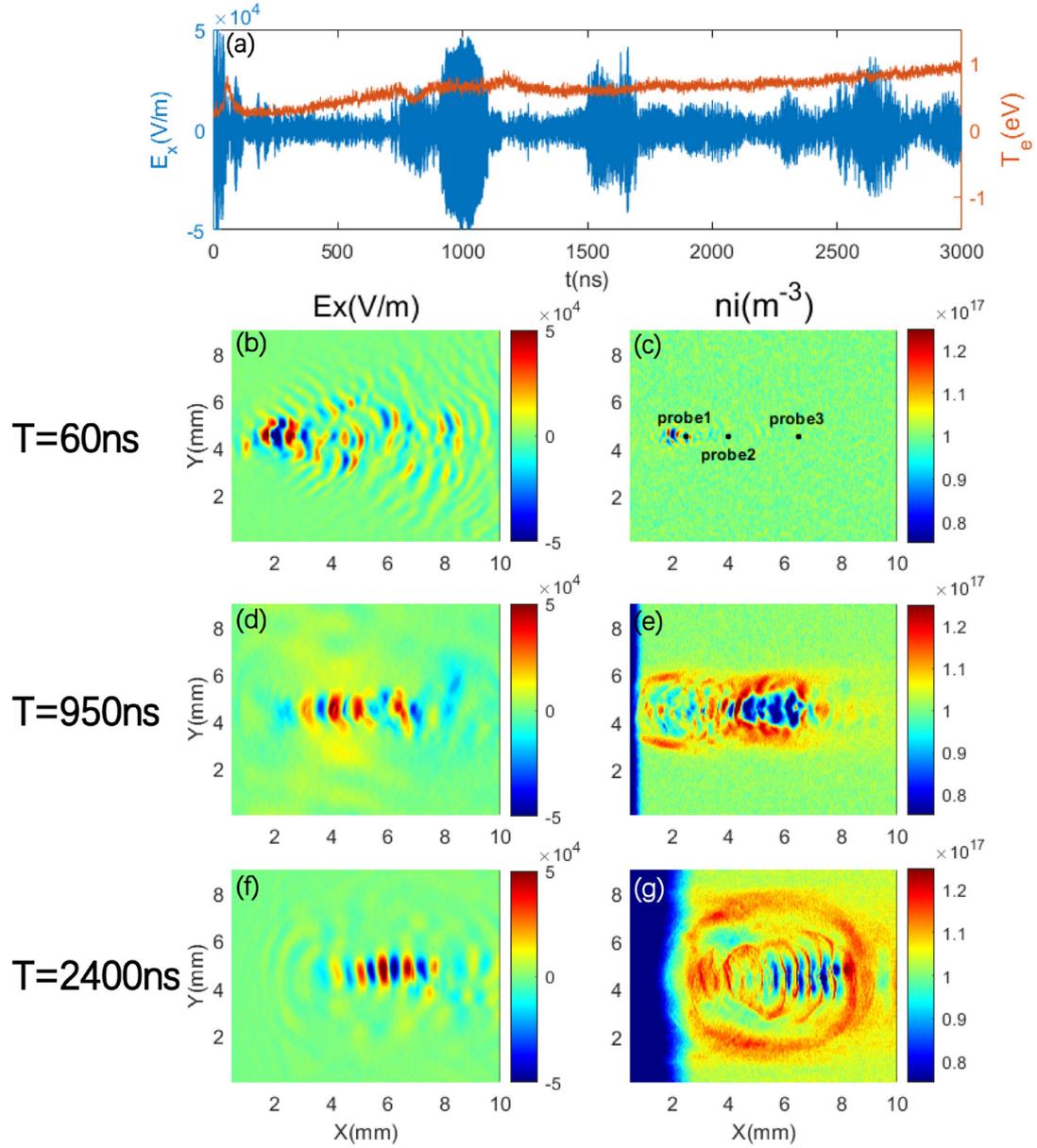

**Figure 11**: Evolution of turbulence for Case 1 with a narrower beam. The beam now is injected only in the spatial interval $y = 3.9 - 5.1 mm$. Blue line in (a) shows the probe diagnostics of $E_x$ (which is the sum of the three probes shown in (c)). Red line in (a) is the electron temperature. (b)-(g) give the snapshots of $E_x$ and $n_i$ at $t = 60ns, 950ns, 2400ns$.

## 4. Theory for Strong Langmuir Turbulence and explanations of simulation results

The nonlinear evolution of the Strong Langmuir Turbulence is very hard to describe comprehensively with an analytical theory. However, it is possible to obtain the criterion for the onset and the initial growth rate of different instabilities in the beam and plasma parameter space analytically by taking into account wave-wave interaction processes. We emphasize that the fluid theory provided below is used *only* to determine the onset of different regimes. A full kinetic analytical approach to this problem will be left for our future investigation. We employ the two-fluid equations for the bulk plasma [81]:

$$\frac{\partial n_e}{\partial t} + \frac{\partial}{\partial x}(n_e \boldsymbol{u}_e) = 0, \tag{1}$$

$$\frac{\partial n_i}{\partial t} + \frac{\partial}{\partial x}(n_i \boldsymbol{u}_i) = 0, \tag{2}$$

$$n_e m \left(\frac{\partial \boldsymbol{u}_e}{\partial t} + \boldsymbol{u}_e \frac{\partial}{\partial x}\boldsymbol{u}_e + \nu_e \boldsymbol{u}_e\right) = -\gamma_e T_e \frac{\partial}{\partial x} n_e - n_e e \boldsymbol{E}, \tag{3}$$

$$n_i M \left(\frac{\partial \boldsymbol{u}_i}{\partial t} + \boldsymbol{u}_i \frac{\partial}{\partial x}\boldsymbol{u}_i + \nu_i \boldsymbol{u}_i\right) = -\gamma_i T_i \frac{\partial}{\partial x} n_i + n_i e \boldsymbol{E}, \tag{4}$$

$$\frac{\partial}{\partial x}\boldsymbol{E} = \frac{e}{\epsilon_0}(n_i - n_e - n_{be}), \tag{5}$$

$$\epsilon_0 \frac{\partial \boldsymbol{E}}{\partial t} = -e(n_i \boldsymbol{u}_i - n_e \boldsymbol{u}_e - \boldsymbol{j}_{be}), \tag{6}$$

where $m$ and $M$ are the mass of electrons and ions, $\nu_e$ and $\nu_i$ are collisional frequency with neutrons for electrons and ions, $\gamma_e$ and $\gamma_i \approx 1$ are adiabatic index. As we are considering a fluid model, closures are needed to treat the pressure term, most importantly, $\gamma_e$ self-consistently and kinetically. Here, we take $\gamma_e \approx 3$ for $\omega \gg v_{the}k$, and $\gamma_e \approx 1$ for $\omega < v_{the}k$. For different frequencies and phase velocities, $\gamma_e$ would take different values to make the corresponding results be consistent with the dispersion relations calculated using kinetic approach. This treatment is commonly used by previous authors [75,76,82,83]. Taking $\gamma_e \approx 1$ for low-frequency motions corresponds to the Boltzmann equilibrium, which is quite straightforward to understand. A qualitative justification for taking $\gamma_e \approx 3$ at high frequency, not fully addressed in prior work, is given in Appendix 2. $n_{be}$ and $\boldsymbol{j}_{be}$ are the beam density and beam current, respectively. $n_e$ and $\boldsymbol{u}_e$ are the density and velocity of bulk electrons only.

Now we apply the multiple scale-separation approach [84] and separate the electric field and electron quantities into a high-frequency part and low-frequency part (the ion equations only evolve on low frequency and are therefore not split).

$$E = E_l + E_h, \tag{7}$$

$$n_e = n_l + n_h, \tag{8}$$

$$\boldsymbol{u}_e = \boldsymbol{u}_l + \boldsymbol{u}_h, \tag{9}$$

where $g_l$ ($g = n, E, \boldsymbol{u}$) is defined as $g_l \doteq \overline{g(\boldsymbol{x},t)} = 1/T \int_{t-T/2}^{t+T/2} g(\boldsymbol{x},t)dt$ and $T$ is chosen such that $\omega_{IAW}^{-1} \gg T \gg \omega_{pe}^{-1}$, (since in our simulations, $\omega_{pe} \sim 10^{10} s^{-1}$, $\omega_{IAW} \sim 10^7 s^{-1}$) which is much slower comparing with the electron plasma wave dynamics, but much faster comparing with the ion dynamics. Therefore, $g_h \doteq g - g_l$. Eq. (1)-(6) could therefore be split into a high-frequency part and a low-frequency one. After tedious but straightforward derivations shown in Appendix 1, we obtain the three most important equations, which constitute the modified Zakharov equations:

$$\left(2i\omega_0 \frac{\partial}{\partial t} + 2i\omega_0 \Gamma_e \frac{n_l}{n_0} + \omega_0^2 - \frac{n_l e^2}{\epsilon_0 m} + \frac{3T_e}{m} \frac{\partial^2}{\partial x^2}\right)\widetilde{\boldsymbol{E}}_h = \widetilde{\boldsymbol{S}}, \tag{10}$$

$$\frac{\gamma_e T_e}{m} \frac{\partial^2}{\partial x^2} \ln n_l - \frac{e^2}{\epsilon_0 m}(n_{bl} + n_l - n_i) = -\frac{\epsilon_0}{4n_0 m} \frac{\partial^2}{\partial x^2}\left|\widetilde{\boldsymbol{E}}_h\right|^2. \tag{11}$$

$$\left(\frac{\partial^2}{\partial t^2} + \Gamma_i \frac{\partial}{\partial t} - \frac{\gamma_i T_i}{M} \frac{\partial^2}{\partial x^2}\right)\frac{n_i}{n_0} - \frac{T_e}{M} \frac{n_i}{n_l} \frac{\partial^2}{\partial x^2} \frac{n_l}{n_0} = \frac{\epsilon_0}{4n_0 M} \frac{\partial^2}{\partial x^2}\left|\widetilde{\boldsymbol{E}}_h\right|^2 \tag{12}$$

Eq. (10) describes the evolution of the electric field envelope, where $\widetilde{S}$ is a source term related to the kinetic beam, which creates the initial saturated wave envelope $|\widetilde{\boldsymbol{E}}_{h0}|$, $\boldsymbol{E}_h(x,t) = \frac{1}{2}[\widetilde{\boldsymbol{E}}_h(x,t)e^{-i\omega_0 t} + \widetilde{\boldsymbol{E}}_h^\star(x,t)e^{i\omega_0 t}]$. The expression of this source term for our model is given in Appendix 1. $\Gamma_e$ is the damping rate of the wave electric field. Note that the source term here is implemented just for a closure. When comparing with simulations, we estimated the saturation level of $|\widetilde{\boldsymbol{E}}_{h0}|$ due to wave trapping using Eq. (24) below rather than solving Eq. (10) directly. This initial saturated wave envelope would be further subjected for wave-wave nonlinear processes, whose initial stage could be described by our two fluid model. Our main focus below is to consider the instability of a perturbation $\widetilde{\boldsymbol{E}}_{h1}$ to this given wave envelope $\widetilde{\boldsymbol{E}}_{h0}$ and solve for the threshold and initial growth rate. Eq. (12) describes evolution of ion density, which is determined by the interaction with ponderomotive force. $\Gamma_i$ is the damping rate of ion acoustic waves, whose expression will be given later. Eq. (11) is the equation for low-frequency electron density perturbation. Eq. (11) describes balance between the ponderomotive force (right hand side, the third term in the equation) and the pressure force of the bulk electrons (the first term on LHS) and the electrostatic force resulting from charge separation which tends to drag the electrons back to a quasi-neutral state (the second term on LHS). It can be clearly seen that the low frequency motion of electrons is balanced by three terms. Depending on magnitude of different terms, Eq. (11) describes three different cases:

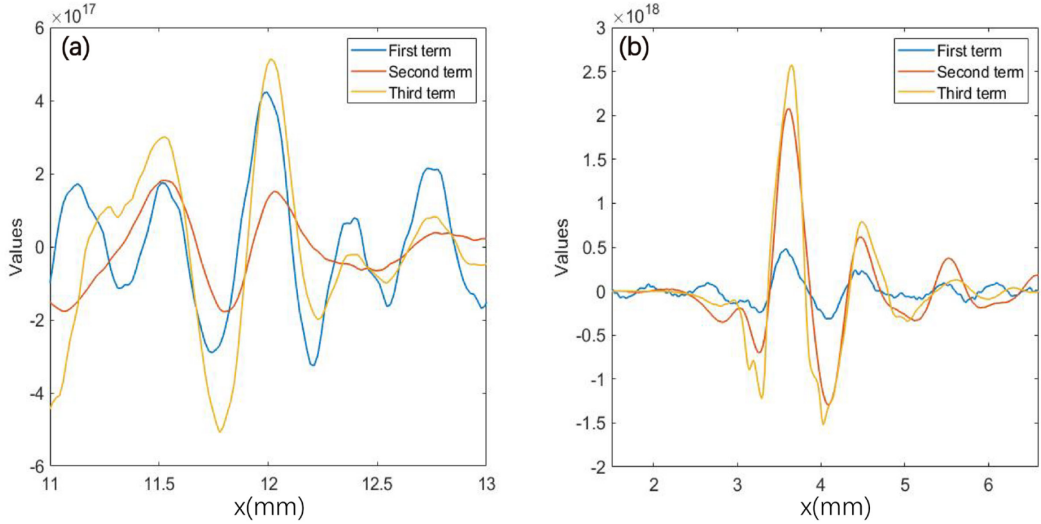

**Figure 12**: Different terms of Eq. (11) at initial instability growth for different cases: (a) verifies that the first term in LHS (blue line) of Eq. (11) nearly balances the term in RHS (yellow line) for Case 2. The data is taken for Case 2 at $t = 128.4ns$ and $y = 4.2mm$. The time average is taken over the neighboring $15.1ns$ (50 electron plasma period). (b) shows that the second term in LHS (red line) of Eq. (11) nearly balances the term in RHS (yellow line) for Case 1. The data is taken for Case 1 at $t = 16.1ns$ and $y = 6.8mm$. The time average is taken over the neighboring $3.0ns$ (10 electron plasma period).

## 4.1 SWMI Regime

**In this regime the instability evolves sufficiently slow so that the charge-neutrality condition, $n_{bl} + n_l - n_i \approx 0$ approximately holds. The first term nearly balances the third term in Eq. (11), as is shown in Fig. 12.**

Substituting the simplified Eq. (11) into Eq. (10) and after some derivations shown in Appendix 1, we obtain the equation for the perturbation $\widetilde{E}_{h1}$ to the initial saturated wave packet:

$$\left(2i\omega_{pe}\frac{\partial}{\partial t} + 2i\omega_{pe}\Gamma_e + \frac{\omega_{pe}^2 \epsilon_0 |\widetilde{E}_{h0}|^2}{2n_0 T_e} + \frac{3T_e}{m}\frac{\partial^2}{\partial x^2}\right)\widetilde{E}_{h1} + \frac{\omega_{pe}^2 \epsilon_0 \widetilde{E}_{h0}^2}{4n_0 T_e}\widetilde{E}_{h1}^* = 0, \quad (13)$$

from which, we could obtain the threshold and growth rate of SWMI by taking $\widetilde{E}_{h1} \sim e^{ikx + \gamma_{SWMI} t}$:

$$\frac{\epsilon_0 |\widetilde{E}_{threshold,SWMI}|^2}{4n_0 T_e} = max\left[(k\lambda_{De})^2, \frac{2\Gamma_e}{\omega_{pe}}\right], \quad (14)$$

$$-3\left(\frac{\epsilon_0 |\widetilde{E}_{h0}|^2}{4n_0 T_e}\right)^2 + 12k^2\lambda_{De}^2 \frac{\epsilon_0 |\widetilde{E}_{h0}|^2}{4n_0 T_e} - 9k^4\lambda_{De}^4 - \left(2\frac{\Gamma_e}{\omega_{pe}} + 2\frac{\gamma_{SWMI}}{\omega_{pe}}\right)^2 = 0. \quad (15)$$

Here $|\widetilde{E}_{h0}|$ is the amplitude of the initial saturated electric field envelope and we take $k \sim k_0 = \omega_{pe}/v_b$ since what we need to consider is the stability of the initial wave packet created by the beam. The criterion (14) implies from physical point of view that the electric field must be strong enough to overcome two counteracting effects: the damping and the dispersion of plasma waves in the nonlinear evolution must be both small enough so that the wave could grow locally concentrating into smaller and smaller regions. From mathematical point of view, since wave-vector $k$ is not a free parameter for our system here (initial value of $k$ is about $k_0$), the combination of two components in Eq. (14) gives a *sufficient condition* for Eq. (15) to have real solution if growth rate is set to be zero. In other words, as long as electric field is greater than the one given by Eq. (14), the instability would grow. This is why the threshold has two components. Above this threshold, large amplitude standing Langmuir waves begin to grow locally and modulate the beam-created wave packet (see Case 2), which is why we believe this threshold determines the boundary of a Strong Turbulent Regime. Because a strong standing wave is generated in this process, we, therefore, call it "Standing Wave Modulational Instability" (SWMI). Comparing with the threshold for Langmuir collapse proposed by V. E. Zakharov [85], we also take into account wave damping in the second term of the threshold in Eq. (14). Eq. (15) here is very similar to results in Bellan's book, Eq. (15.149) [75]. The small difference is because Bellan used, $n_l \approx n_i$ whereas we adopted $n_l \approx n_i - n_{bl}$. Comparing with the previous work by Papadopoulos [86], our result here is not limited to *small amplitude perturbations* $\delta n_e \ll n_{p0}$ (meaning the Langmuir wave amplitude has to be small). We also do not assume spatially homogeneous plane wave (i.e., we retained the spatial derivative term in the derivations of Eq. (14) and Eq. (15)). Note that the instability given by SWMI is still on the ion time scale, namely, $\gamma_{SWMI} \leq \omega_{IAW}$ since the charge neutrality condition is approximately satisfied.

The SWMI is typically saturated by the following two mechanisms: First, as the wave energy is concentrated into a smaller and smaller region in the nonlinear stage, the wave number $k$ gradually increases, giving rise to also stronger damping of the waves (transit-time damping). The threshold of SWMI therefore increases, which could inhibit the growth of SWMI. Second, as is evident from the simulation results, if ion density perturbations are strong, the beam decouples from the previously excited plasma waves and therefore stopping the wave generation and instability growth. The quantitative analysis of the SWMI saturation is beyond the scope of this paper.

**4.2 EMI Regime**

**Electron density changes much faster than ions, ions don't have time to response. The electrostatic force balances the ponderomotive force, namely, the second term nearly balances the third term in Eq. (11), which is shown in Fig. 12.**

After tedious but straightforward derivations in Appendix, we obtain an equation for estimating $\widetilde{E}_{h1}$:

$$\left(2i\omega_{pe}\frac{\partial}{\partial t} + 2i\omega_{pe}\Gamma_e\right)\widetilde{E}_{h1} - \frac{\epsilon_0}{4m_e n_0}\widetilde{E}_{h1}\frac{\partial^2}{\partial x^2}\left|\widetilde{E}_{h0}\right|^2 - \frac{\epsilon_0}{4m_e n_0}\widetilde{E}_{h0}\frac{\partial^2}{\partial x^2}(\widetilde{E}_{h0}^*\widetilde{E}_{h1} + \widetilde{E}_{h0}\widetilde{E}_{h1}^*) + \left(\frac{n_b}{n_0}\omega_{pe}^2 + \frac{3T_e}{m}\frac{\partial^2}{\partial x^2}\right)\widetilde{E}_{h1} = 0 \quad (16)$$

From this equation, the threshold of EMI could be expressed by:

$$\frac{\epsilon_0|\tilde{E}_{threshold,EMI}|^2}{4n_0 T_e} = max\left[1 - \frac{n_b}{3n_0}\frac{1}{k^2\lambda_{De}^2}, 2\frac{\Gamma_e}{\omega_{pe}}\frac{1}{k^2\lambda_{De}^2}\right]. \quad (17)$$

Depending on the beam density, the growth rate is given by different relations:

If $n_b/n_0 < 3k^2\lambda_{De}^2$, the growth rate is given by:

$$-3\left(k^2\lambda_{De}^2\frac{\epsilon_0|\tilde{E}_{h0}|^2}{4n_0 T_e}\right)^2 + 4\left(3k^2\lambda_{De}^2 - \frac{n_b}{n_0}\right)\left(k^2\lambda_{De}^2\frac{\epsilon_0|\tilde{E}_{h0}|^2}{4n_0 T_e}\right) - \left(3k^2\lambda_{De}^2 - \frac{n_b}{n_0}\right)^2 - \left(2\frac{\Gamma_e}{\omega_{pe}} + 2\frac{\gamma_{EMI}}{\omega_{pe}}\right)^2 = 0. \quad (18)$$

If $n_b/n_0 > 3k^2\lambda_{De}^2$, the growth rate becomes:

$$2\frac{\gamma_{EMI}}{\omega_{pe}} = \frac{n_b}{n_0} + k^2\lambda_{De}^2\frac{\epsilon_0}{4n_0 T_e}|\tilde{E}_{h0}|^2 - 3k^2\lambda_{De}^2 - 2\frac{\Gamma_e}{\omega_{pe}}, \quad (19)$$

where we adopted $k \sim k_0$, $n_b$ is the beam density. Similar to Eq. (14), Eq. (17) implies that the electric field must be sufficiently strong to modify the electron dynamics before ions respond to strong electric field. At the same time, the damping must be small enough so that the wave could grow locally. Above this threshold, the strong ponderomotive force of the Langmuir wave packet causes a breakdown of charge neutrality condition. It is the electrostatic force resulting from charge separation that nearly balances the ponderomotive force (see also Appendix). We, therefore, believe it is this instability that gives the strong localization of Langmuir waves in Case 1. Because this instability in the initial stage evolves electrons only, we call it "Electron Modulational Instability" (EMI). Note that EMI is essentially different from both the classical modulational instability and parametric decay instability because it does not involve ion dynamics in its initial stage and is much faster than the classical Langmuir collapse process.

The two nonlinear saturation mechanisms for EMI is similar to SWMI. However, the saturation amplitude in EMI is much stronger than that in SWMI for the following two reasons: First, the saturation amplitude of electric field before wave-wave instability is stronger in EMI due to the large density and energy of the beam (see Eq. (24)); Second, in EMI, the localized field does not need to wait for the ion response like that in SWMI, therefore the ions do not limit the growth of field in the initial stage. After the ions are slightly perturbed, the field could be even more localized, giving rise to an even stronger field. Such a strong localized field in EMI could explain the rapid beam scattering and efficient electron heating observed in simulations.

**4.3 PDI Regime**

**Charge neutrality condition, $n_l = n_i$ is valid at all times. Eq. (10) therefore needs to be coupled with ion equation to solve the system self-consistently. This case is most commonly studied by previous authors.**

The equation for perturbation $\widetilde{E}_{h1}$ and associated ion equation read:

$$\left(2i\omega_{pe}\frac{\partial}{\partial t} + 2i\omega_{pe}\Gamma_e + \frac{3T_e}{m}\frac{\partial^2}{\partial x^2}\right)\widetilde{E}_{h1} = \omega_{pe}^2 \frac{\delta n_i}{n_0}\widetilde{E}_{h0}, \tag{20}$$

$$\left(\frac{\partial^2}{\partial t^2} + \Gamma_i\frac{\partial}{\partial t} - c_S^2\frac{\partial^2}{\partial x^2}\right)\frac{\delta n_i}{n_0} = \frac{\epsilon_0}{4Mn_0}\frac{\partial^2}{\partial x^2}\left(\widetilde{E}_{h0}^*\widetilde{E}_{h1} + \widetilde{E}_{h0}\widetilde{E}_{h1}^*\right), \tag{21}$$

which could be further simplified to give solutions for PDI. Because the PDI is not the main focus of this paper, we only present a simplified form of threshold of PDI calculated by Nishikawa [58,59], the expression is:

$$\frac{\epsilon_0|\widetilde{E}_{threshold,PDI}|^2}{4n_0T_e} = 4\frac{\Gamma_e}{\omega_{pe}}\frac{\Gamma_i}{\omega_{IAW}}, \tag{22}$$

where $\Gamma_i \approx \sqrt{\pi/8}\,\omega_{IAW}/(1 + k_{IAW}^2\lambda_{De}^2)^{3/2}[\sqrt{m_e/m_i} + (T_e/T_i)^{3/2}\exp(-1.5 - T_e/2T_i/(1 + k_{IAW}^2\lambda_{De}^2))] + \nu_{in}/2$, $\omega_{IAW}^2 \approx k_{IAW}^2 c_S^2/(1 + k_{IAW}^2\lambda_{De}^2)$ [75]. The growth rate reads:

$$\frac{\gamma_{PDI}}{\omega_{pe}} = \frac{1}{\Gamma_e + \Gamma_i}\left(\frac{T_e k_{IAW}^2}{16m_i\omega_{IAW}}\frac{\epsilon_0 E^2}{n_0 T_e} - \frac{\Gamma_e\Gamma_i}{\omega_{pe}}\right). \tag{23}$$

There are two effects contributing to the saturation of PDI. First, the heating of electrons increases the damping rate of Langmuir waves and ion acoustic waves, which results in a higher threshold of PDI. Second, the generated backward Langmuir waves and forward ion acoustic waves propagate out of the source region, transferring energy out of it. A saturation is reached when the energy input balances to the energy transferred out [87].

To obtain a threshold for different regimes in the parameter space, we need to combine the thresholds above with an expression for the saturated electric field before wave-wave processes develop. In the simulations, it is the beam relaxation mechanism that determines the initial saturated electric field, $E_{sat}$ in the simulation domain if there is no strong turbulence. In our simulations, the QL approach breaks down because we consider an intense and mono-energetic beam, as is often used in practical experimental applications. A standard approach as, e.g., in Matsiborko et *al.* [88] described the correct beam relaxation mechanism in the cold beam limit using momentum conservation. The saturation mechanism is the wave trapping rather than the QL relaxation. The saturated electric field can be estimated by:

$$\frac{\epsilon_0 E_{sat}^2}{4 n_0 T_e} = \frac{9}{8}\left(\frac{n_b}{n_0}\right)^{4/3}\frac{m_e v_b^2}{2T_e}, \qquad (24)$$

where $n_b$ is beam density and $v_b$ is beam velocity. Eq. (24) has been verified many times in experimental studies, see e.g., [38] and simulations [54,56]. We therefore obtain the threshold for a Strong Turbulent Regime onset by taking $E_{sat} > E_{threshold,SWMI}$:

$$\frac{9}{8}\left(\frac{n_b}{n_0}\right)^{4/3}\frac{m_e v_b^2}{2T_e} > max\left[\frac{2\Gamma_e}{\omega_{pe}},(k\lambda_{De})^2\right]. \qquad (25)$$

The boundary of EMI onset:

$$\frac{9}{8}\frac{m_e v_b^2}{2T_e}\left(\frac{n_b}{n_0}\right)^{4/3} > max\left[1 - \frac{n_b}{3n_0}\frac{1}{k^2\lambda_{De}^2}, 2\frac{\Gamma_e}{\omega_{pe}}\frac{1}{k^2\lambda_{De}^2}\right], \qquad (26)$$

and the boundary of PDI:

$$\frac{9}{8}\frac{m_e v_b^2}{2T_e}\left(\frac{n_b}{n_0}\right)^{4/3} > 4\frac{\Gamma_e}{\omega_{pe}}\frac{\Gamma_i}{\omega_{IAW}}. \qquad (27)$$

**4.4 Verification with simulations**

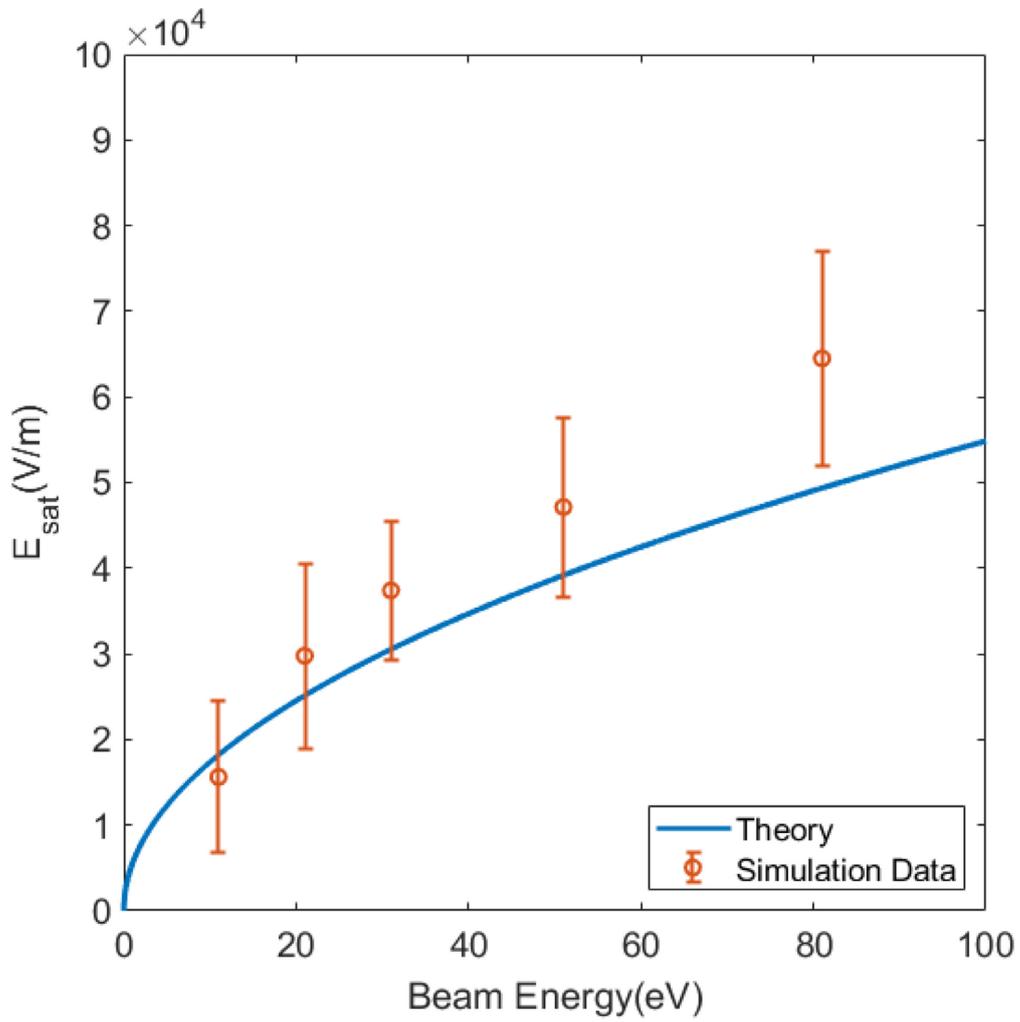

**Figure 13**: The saturated electric field before onset of the modulational instability as a function of beam energy $E_b$ for $n_b/n_p = 0.015$; it shows applicability of Eq. (24) in estimating the initial saturated wave packet amplitude before the onset of modulational instability. The blue curve shows theoretical value given by Eq. (24) and the red dots show the simulation data before SLT develops. The error bars are obtained by taking data from several different snapshots over one or several cases.

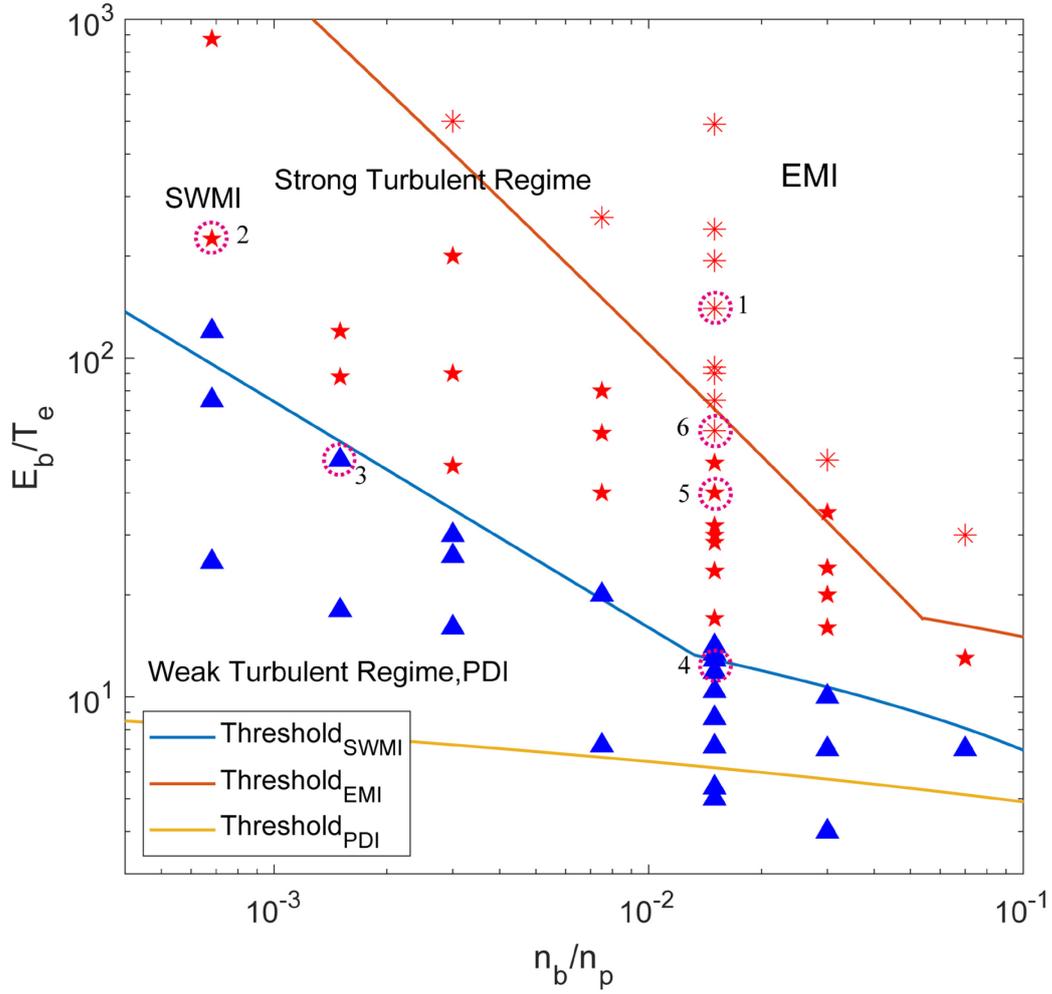

**Figure 14**: Parameter space of ratio of the beam energy to the bulk electron temperature $E_b/T_e$ versus the ratio of the beam density to the plasma density, $n_b/n_p$. The blue line shows the threshold Eq. (25) and the red line shows the threshold Eq. (26), respectively. The yellow curve shows the threshold of Langmuir PDI given by Eq. (27). Red and blue markers show the cases with and without strong turbulence, respectively. Red plus-over-an-x markers show the cases with EMI. Pink circles mark different cases used for detailed analysis in this paper. The numbers denote the Case number shown in Table 2. In total 57 simulations are shown here (all the simulation parameters are given in the table in Appendix 3).

To further verify the theory, we performed 57 simulations varying the beam energy and initial plasma electron temperature (all parameters are shown in Table A1). Fig. 13 verified the correctness of Eq. (24) in estimating the saturated electric field before the onset of strong turbulence. This verifies that the saturation mechanism of beam-plasma interaction is due to the wave trapping of beam particles rather than the formation of the plateau of the EVDF as assumed by the QL theory. Before the onset of strong turbulence, the electron distribution function is approximately Maxwellian, the damping for the wave is given by [75]:

$$\Gamma_e \approx \sqrt{\pi/8}\omega_{pe}/(k\lambda_{De})^3 \exp(-1.5 - 1/2/(k\lambda_{De})^2) + \nu_{en}/2, \qquad (28)$$

where $\nu_{en} \approx n_n v_b \sigma_{en,elas}$ is the collisional frequency between electrons and neutrals, where $n_n$ is the density of neutral atoms, $v_b$ is the beam velocity, $\sigma_{en,elas}$ is elastic collision cross-section fitted from the data available [89,90]. Here the wave number $k$ is also taken to be comparable to $k_0$. For nonlinear evolution of SLT, several phenomenological transit-time damping models could be used in the place of linear Landau damping [40-43]. Here, we use the linear damping model in the analytical theory because it is used only at the beginning of the modulational instabilities, when transit-time damping is not strong. In the simulations, we calculated the damping rate at the moment before the strong turbulence occurs (for cases without strong turbulence the time is taken according to the cases where strong turbulence occurs). Fig. 14 shows that Eq. (25) provides a reasonable estimation of whether strong turbulence occurs or not under a large parameter regime with different beam densities (two orders of magnitude). In the simulations, each case must satisfy the following three criteria to be qualified as the one with the strong turbulence (red markers in the figure):

1. The three stages must be seen in the energy transfer plot.
2. There is a clear standing wave feature and the associated ion density depletion. The intensity of the standing wave must go beyond the saturation value given by Eq. (24).
3. Such a process must occur faster compared with PDI, for SWMI to dominate over PDI.

We also observed charge separation and wave localization faster than the ion frequency (indicating EMI) for 12 cases above the threshold given by Eq. (26)

(including Case 1, denoted by red plus-over-an-x markers), confirming the reliability of Eq. (26). From Fig. 14, we see that the threshold of PDI given by Eq. (27) is low compared with SWMI and EMI. We define the regime where SWMI does not occur while PDI still occurs as "Weak Turbulent Regime", which is shown in the parameter space by the region between the boundary of Strong Turbulent Regime and the yellow line in Fig. 14. Please note that in plotting the threshold of PDI (the yellow line) we take a fixed $T_e/T_i = 20$ in order to reduce the expression to a scaling of $n_b/n_0$ and $E_b/T_e$.

When the Landau damping can be neglected in Eq. (25) and (26), namely, when the beam is very energetic and wavelength is long, the criterion can be well approximated by the following two scalings:

$$\frac{E_b}{T_e} \sim \frac{2}{3}\left(\frac{n_b}{n_0}\right)^{-\frac{2}{3}}, \tag{29}$$

$$\frac{E_b}{T_e} \sim \left(\frac{9}{8}\left(\frac{n_b}{n_0}\right)^{\frac{4}{3}} + \frac{2}{3}\frac{n_b}{n_0}\right)^{-1}, \tag{30}$$

which would follow the shape of the left branch of blue line and red line shown in Fig. 14. The scaling given by Eq. (29) is different from the one given by Galeev $E_b/T_e \sim (n_b/n_0)^{-1/3}$ in 1977 [91] because the authors assumed quasilinear relaxation of the beam while in our case the saturation mechanism is wave trapping. In principle, quasilinear relaxation is only valid when $\Delta v_b/v_b > (n_b/n_0)^{1/3}$, $\Delta v_b$ is the velocity spread of the beam, which is comparable to beam thermal velocity. A typical electron beam in the beam discharge experiments, however, is usually very cold: $T_{eb} \approx 0.2eV$ and $\Delta v_b/v_b \ll (n_b/n_0)^{1/3}$ [6,38], which means that quasilinear theory rarely holds. The new scaling law was given by Eq. (30) also separates a new regime that has not been studied in detail by other people, to the best of our knowledge. The Langmuir wave localization is more rapid than previously was thought in this regime.

## 5. Conclusions

In this paper, we extensively studied the collective processes of electron beam-plasma interaction under different parameters using high resolution 2D PIC simulations and analytical theories. We showed that this process, in terms of wave-wave nonlinear interaction, can be classified into four different regimes. We identified a new regime where the Langmuir wave packet could grow locally faster than the ion frequency. The dominant instability in this regime is called Electron Modulation Instability, EMI. The EMI regime exhibits a repetitive (intermittent) three-stage process and is often associated with strong energy transfer from the wave electric field to bulk electrons, strong beam scattering, and a broad electric field energy spectrum. The duration of the intermittent feature depends strongly on the system size and beam width (see Section 3.5.2). We showed that the boundary of the Strong Turbulent Regime is determined by the threshold of SWMI. The large density perturbations resulting from the ponderomotive force of the intense standing Langmuir waves in a Strong Turbulent Regime causes the beam to decouple from the plasma. The Langmuir waves in the Weak Turbulent Regime are dominated by the PDI. Four cases in the three regimes are shown and compared in detail. Several interesting phenomena are briefly discussed. In the theory part, using multi-fluid equations, we derived an analytical model determining the boundary and estimate for the initial growth rate in different regimes, which is verified via the 57 simulation cases.

For the first time, we propose a comprehensive analytical criterion or scaling law for the onset of Strong Langmuir Turbulence determining the boundary of different regimes in the parameter space. This can guide past and future numerical and experimental studies of beam-plasma interactions such as in pulsed-beam plasma devices and in space plasmas.

We show that the strong Langmuir turbulence exhibits rich physics which hasn't yet been fully understood, such as the formation mechanism of strong perpendicular electric fields (giving rise to the beam scattering), intermittent feature, the mechanism of the kappa distribution function formation, and the k-spectrum (especially a $-5$ spectrum). Detailed analyses of these physical features are left for future work.

In this work, we limited ourselves to 2D simulations without a background magnetic field. A 3D PIC simulation of Langmuir turbulence and extension to electromagnetic case are left for future studies. The analytical model proposed here is based upon the fluid assumption, which is perhaps valid only for predicting the onset of different modulational instabilities and may not be valid for describing strong kinetic effects in the nonlinear stage of the modulational instabilities. A full kinetic approach to this problem is left for future investigations.

## Acknowledgements

The authors thank the referees for the careful reading of the manuscript and very helpful comments that helped us improve the manuscript. We thank Prof. Ilya Dodin, Prof. Quanming Lu, Prof. Chuanbing Wang, Dr. Stephan Brunner, Dr. Justin Ball, Dr. Sarveshwar Sharma and Dr. Liang Xu for the fruitful discussions. The work of I.K., A.K, and D. S. was supported by the Princeton Collaborative Research Facility (PCRF) and Laboratory Directed Research & Development (LDRD) projects, which are funded by the U. S. Department of Energy (DOE) under Contract No. DE-AC02-09CH11466.

## Appendix 1: Modified Zakharov equations

### Introduction and motivation

It has been 50 years since the well-known set of Zakharov equations was introduced in 1972. It is a nonlinear Schrodinger equation coupled with an ion-acoustic equation describing the evolution of the electric field envelope [31]. Since then, several important works, such as Ref. [2-5], came out in which Zakharov equations were applied in various contexts. One essential assumption in deriving these equations is the charge quasi-neutrality: $n_e \approx n_i$. However, such condition is only valid when the system evolves sufficiently slow, roughly on the ion acoustic time scale, meaning that the classical Zakharov equations could be used. Therefore, the equations should be modified in order to study the case when an instability develops faster than the ions respond and the charge quasi-neutrality condition does not hold.

The purpose of this Appendix is to revisit the derivation of nonlinear wave-coupling equations staring with a basic two-fluid description of the plasma and employing a time-scale separation approach. We make a point to clarify the underlying assumptions, as well as their physical meaning. Our numerical simulations will be used to guide and illustrate the process. We show that depending on certain assumptions, there are three different ways to obtain the wave coupling equations.

We show that, quite remarkably, there is a new regime dominated by a very fast modulational instability. Namely, the wave packet grows locally faster than ion time scale. We call such an instability an Electron Modulational Instability (EMI). It will be shown how the traditional Zakharov equations are modified in each of the three regimes. We also present a brief discussion on how some assumptions made in the prior work (such as the smallness of the electron density perturbation) are not always valid. Thus, the instability threshold we obtain is different from previous results.

**Derivation of reduced equations**

The situation we consider, as seen in the presented simulation cases, is when the Langmuir wave packet is excited by an electron beam in the plasma. Our purpose is to consider the nonlinear wave-wave interaction. We assume that the wave packet already exists, taking the form of $\boldsymbol{E} = \boldsymbol{E}_0 + \boldsymbol{E}_1$, where $\boldsymbol{E}_0$ defines the shape of the packet and $\boldsymbol{E}_1 \ll \boldsymbol{E}_0$ is a small perturbation. We then consider the interaction of such a wave packet with the bulk plasma populations. It is noted that, for the purpose of interpreting our simulations, the analytical theory below applies *only* at the early stage of the instability development (otherwise the assumptions underlying the fluid model cease to be valid).

First, we state the following assumptions adopted to derive the fluid equations:
1. Fluid assumption: $k_0^2 \lambda_{De}^2 \ll 1$.

2. One-dimensional variation in space: the interaction is described as a one-dimensional process, with both the beam and the waves propagating in $x$ direction. A two-fluid description for the bulk plasma is adopted:

$$\frac{\partial n_e}{\partial t} + \frac{\partial}{\partial x}(n_e \boldsymbol{u}_e) = 0, \tag{1}$$

$$\frac{\partial n_i}{\partial t} + \frac{\partial}{\partial x}(n_i \boldsymbol{u}_i) = 0, \tag{2}$$

$$n_e m \left(\frac{\partial \boldsymbol{u}_e}{\partial t} + \boldsymbol{u}_e \frac{\partial}{\partial x} \boldsymbol{u}_e + \nu_e \boldsymbol{u}_e\right) = -\gamma_e T_e \frac{\partial}{\partial x} n_e - n_e e \boldsymbol{E}, \tag{3}$$

$$n_i M \left(\frac{\partial \boldsymbol{u}_i}{\partial t} + \boldsymbol{u}_i \frac{\partial}{\partial x} \boldsymbol{u}_i + \nu_i \boldsymbol{u}_i\right) = -\gamma_i T_i \frac{\partial}{\partial x} n_i + n_i e \boldsymbol{E}, \tag{4}$$

$$\frac{\partial}{\partial x} \boldsymbol{E} = \frac{e}{\epsilon_0}(n_i - n_e - n_{be}), \tag{5}$$

$$\epsilon_0 \frac{\partial \boldsymbol{E}}{\partial t} = -e(n_i \boldsymbol{u}_i - n_e \boldsymbol{u}_e - \boldsymbol{j}_{be}), \tag{6}$$

where $m$ and $M$ are the masses of electrons and ions, $\nu_e$ and $\nu_i$ are the respective frequencies of collisions with neutrals, $\gamma_e$ and $\gamma_i \approx 1$ are adiabatic indices, $n_e$ and $\boldsymbol{u}_e$ describe the bulk electrons only, while $n_{be}$ and $\boldsymbol{j}_{be}$ are the beam density and current, respectively, whose expressions are given by Eq. (8). For the fluid model to be adequate, it is necessary that the pressure term represents a correct kinetic closure. In our case, for electrons this is achieved by setting $\gamma_e \approx 3$ for $\omega \gg v_{the} k$ and $\gamma_e \approx 1$ for $\omega < v_{the} k$. This treatment makes the result formally consistent with the wave dispersion relation based on the kinetic approach and is commonly used by previous authors [75,76,82,83]. Taking $\gamma_e \approx 1$ for low-frequency motions corresponds to the Boltzmann equilibrium, and a qualitative justification for taking $\gamma_e \approx 3$ at high frequency, not addressed in prior work, is given in Appendix 2.

We further write down the kinetic Vlasov equations for the electron beam:

$$\frac{\partial f_{be}}{\partial t} + \boldsymbol{v}_{be} \frac{\partial f_{be}}{\partial x} - \frac{e\boldsymbol{E}}{m} \frac{\partial f_{be}}{\partial v_{be}} = 0, \tag{7}$$

$$n_{be} = \int f_{be} d^3 v_{be}, \quad \boldsymbol{j}_{be} = \int f_{be} \boldsymbol{v}_{be} d^3 v_{be} \tag{8}$$

The variables $f_{be}$ and $\boldsymbol{v}_{be}$ are introduced explicitly to represent the distribution function and the velocity of the beam electrons. As the beam density is typically very small, we regard the beam as an external source which sustains the initial wave packet $E_0$. Note that the kinetic treatment of the beam is presented only to provide a closure to the set of equations; namely, to introduce the source term $S$ in what follows. In this Appendix, we do not solve for the initial saturated wave amplitude directly based on the source term $S$. Namely, when comparing with simulations, we estimate the saturation value of the initial wave packet using Eq. (24) in the main text rather than directly solving Eq. (29) below. A complete solution to this set of equations is left for future studies.

Now, we split the electric field, electron density, and electron velocity into a low-frequency and a high-frequency component:

$$\boldsymbol{E} = \boldsymbol{E}_l + \boldsymbol{E}_h, \qquad (9)$$

$$n_e = n_l + n_h, \qquad (10)$$

$$\boldsymbol{u}_e = \boldsymbol{u}_l + \boldsymbol{u}_h, \qquad (11)$$

$$f_{be} = f_{bl} + f_{bh}, \qquad (12)$$

$$\boldsymbol{v}_{be} = \boldsymbol{v}_{bl} + \boldsymbol{v}_{bh}, \qquad (13)$$

where $g_l$ ($g = n, E, \boldsymbol{u}, f_b, \boldsymbol{v}_b$) is defined as $g_l \doteq \overline{g(x,t)} = 1/T \int_{t-T/2}^{t+T/2} g(x,t) dt$ and *the time − averaging window T* is chosen such that $\omega_{IAW}^{-1} \gg T \gg \omega_{pe}^{-1}$ (in our simulations, $\omega_{pe} \sim 10^{10} s^{-1}$, $\omega_{IAW} \sim 10^7 s^{-1}$), that is, much longer compared to the electron time scale, but much shorter than ion-acoustic time. Within that range, the averaging window can be varied. Doing so will change some of the equations below. Therefore, $g_h \doteq g - g_l$. Under the given definition of time averaging, the ions will only undergo slow motion, which is the reason why we didn't split the ion density into two parts. Note that the characteristic time scale for the high frequency component is always $\sim \omega_{pe}^{-1}$ while the low-frequency time scale is $\sim T$. Up to this point, our approach is similar to that of Morales and Lee [84] except they used fluid beam while we are using kinetic description for the beam, which is more self-consistent. Now we apply time averaging to Eqs. (1)-(8) and obtain the following equations for the low frequency components:

$$\frac{\partial n_l}{\partial t} + \frac{\partial}{\partial x}(n_l \boldsymbol{u}_l + \overline{n_h \boldsymbol{u}_h}) = 0, \qquad (14)$$

$$\frac{\partial n_i}{\partial t} + \frac{\partial}{\partial x}(n_i \boldsymbol{u}_i) = 0, \tag{15}$$

$$m\left(\frac{\partial \boldsymbol{u}_l}{\partial t} + \boldsymbol{u}_l \frac{\partial}{\partial x}\boldsymbol{u}_l + \overline{\boldsymbol{u}_h \frac{\partial}{\partial x}\boldsymbol{u}_h} + \nu_e \boldsymbol{u}_l\right) = -\gamma_e T_e \frac{\partial}{\partial x}\ln n_e - e\boldsymbol{E}_l, \tag{16}$$

$$n_i M\left(\frac{\partial \boldsymbol{u}_i}{\partial t} + \boldsymbol{u}_i \frac{\partial}{\partial x}\boldsymbol{u}_i + \nu_i \boldsymbol{u}_i\right) = -\gamma_i T_i \frac{\partial}{\partial x} n_i + n_i e \boldsymbol{E}_l, \tag{17}$$

$$\frac{\partial}{\partial x}\boldsymbol{E}_l = \frac{e}{\epsilon_0}(n_i - n_l - n_{bl}), \tag{18}$$

$$\frac{\partial \boldsymbol{E}_l}{\partial t} = -\frac{e}{\epsilon_0}(n_i \boldsymbol{u}_i - n_l \boldsymbol{u}_l - \overline{n_h \boldsymbol{u}_h} - \boldsymbol{j}_{bl}), \tag{19}$$

$$\frac{\partial f_{bl}}{\partial t} + \boldsymbol{v}_{bl}\frac{\partial f_{bl}}{\partial x} + \overline{\boldsymbol{v}_{bh}\frac{\partial f_{bh}}{\partial x}} - \frac{e\boldsymbol{E}_l}{m}\frac{\partial f_{bl}}{\partial v_{be}} - \overline{\frac{e\boldsymbol{E}_h}{m}\frac{\partial f_{bh}}{\partial v_{be}}} = 0, \tag{20}$$

$$n_{bl} = \int f_{bl} d^3 \boldsymbol{v}_{be}, \quad \boldsymbol{j}_{bl} = \int (f_{bl}\boldsymbol{v}_{bl} + \overline{f_{bh}\boldsymbol{v}_{bh}}) d^3 \boldsymbol{v}_{be}, \tag{21}$$

where we have used the properties of the average value. We see that ion equations are only slightly modified. Let us first consider the electron equations and the equations for the electric field. Subtracting the low frequency part of Eq. (14), (16), (19), (20), (21) from the governing equations, we obtain equations for the high-frequency (oscillating) components:

$$\frac{\partial n_h}{\partial t} + \frac{\partial}{\partial x}(n_l \boldsymbol{u}_h + n_h \boldsymbol{u}_h - \overline{n_h \boldsymbol{u}_h}) = 0, \tag{22}$$

$$m\left(\frac{\partial \boldsymbol{u}_h}{\partial t} + \boldsymbol{u}_h \frac{\partial}{\partial x}\boldsymbol{u}_h - \overline{\boldsymbol{u}_h \frac{\partial}{\partial x}\boldsymbol{u}_h} + \nu_e \boldsymbol{u}_h\right) = -\gamma_e T_e \frac{\partial}{\partial x}\ln((n_l + n_h)/n_l) - e\boldsymbol{E}_h, \tag{23}$$

$$\frac{\partial \boldsymbol{E}_h}{\partial t} = -\frac{e}{\epsilon_0}(-n_l \boldsymbol{u}_h - n_h \boldsymbol{u}_h + \overline{n_h \boldsymbol{u}_h} - \boldsymbol{j}_{bh}), \tag{24}$$

$$\frac{\partial f_{bh}}{\partial t} + \boldsymbol{v}_{bh}\frac{\partial f_{bl}}{\partial x} + \boldsymbol{v}_{bl}\frac{\partial f_{bh}}{\partial x} + \boldsymbol{v}_{bh}\frac{\partial f_{bh}}{\partial x} - \overline{\boldsymbol{v}_{bh}\frac{\partial f_{bh}}{\partial x}} - \frac{e\boldsymbol{E}_h}{m}\frac{\partial f_{bl}}{\partial v_{be}} - \frac{e\boldsymbol{E}_l}{m}\frac{\partial f_{bh}}{\partial v_{be}} - \frac{e\boldsymbol{E}_h}{m}\frac{\partial f_{bh}}{\partial v_{be}} +$$

$$\overline{\frac{e\boldsymbol{E}_h}{m}\frac{\partial f_{bh}}{\partial v_{be}}} = 0, \tag{25}$$

$$\boldsymbol{j}_{bh} = \int (f_{bh}\boldsymbol{v}_{bh} - \overline{f_{bh}\boldsymbol{v}_{bh}} + f_{bh}\boldsymbol{v}_{bl} + f_{bl}\boldsymbol{v}_{bh}) d^3 \boldsymbol{v}_{be}. \tag{26}$$

An additional assumption has been made: the terms containing $\boldsymbol{u}_l$ are very small compared to the other terms shown in three equations above. This is reasonable because these equations are describing the high frequency motion of electrons. The time scale is basically on the order of $\omega_{pe}^{-1}$. The low frequency motion of bulk electrons could therefore be neglected.

Next, we consider the high-frequency electric field. Taking a time derivative of Eq. (24) and inserting Eq. (23) into it, under the assumption stated above, we obtain:

$$\frac{\partial^2 E_h}{\partial t^2} = \frac{e}{\epsilon_0}\left(-\frac{n_l \gamma_e T_e}{m}\frac{\partial}{\partial x}\ln\left(1+\frac{n_h}{n_l}\right) - \frac{en_l E_h}{m} - \nu_e n_l u_h\right) + S, \quad (27)$$

where

$$S \doteq \frac{e}{\epsilon_0}\frac{\partial}{\partial t}(n_h u_h - \overline{n_h u_h} + j_{bh}) + \frac{e}{\epsilon_0}n_l\overline{\left(u_h\frac{\partial}{\partial x}u_h - u_h\frac{\partial}{\partial x}u_h\right)}, \quad (28)$$

and we also used the fact that $\partial n_l/\partial t$ is small compared with other terms. Physically, it means that initially, it is mainly the beam current $j_{bh}$ which sustains the initial wave packet $E_0$ because other terms remains to be small in the initial stage. When comparing with simulations, the initial saturation level ($E_0$) can be estimated using Eq. (24) in the main text. The primary focus of this Appendix, however, is the subsequent wave-wave interactions based on this already generated initial wave packet. In Eq. (27), the terms containing $u_h$ and $\gamma_e T_e$ are of a lower order. Now, we combine Eq. (27) with Eq. (23) and the Poisson's equation for $E_h$ to the lowest non-vanishing order (namely, taking $m\frac{\partial u_h}{\partial t} \sim -eE_h$ and $\frac{\partial}{\partial x}E_h = \frac{e}{\epsilon_0}(-n_h)$ and inserting these expressions into Eq. (27)). Then, assuming that $n_h/n_l$ is not very large such that $\ln(1+n_h/n_l) \approx n_h/n_l$ (which is true in most simulation cases), we obtain an equation for $E_h$:

$$\left(\frac{\partial^2}{\partial t^2} + 2\Gamma_e\frac{n_l}{n_0}\frac{\partial}{\partial t} - \frac{\gamma_e T_e}{m}\frac{\partial^2}{\partial x^2}\right)E_h = -\frac{e^2 n_l}{\epsilon_0 m}E_h + S, \quad (29)$$

where we have additionally allowed for the effect of Landau damping: $\Gamma_e = \Gamma_L + \nu_e/2$ is the effective damping rate for the electric field, where $\Gamma_L$ is the Landau damping rate. This approach follows the one employed in previous studies [92,93]. One can actually prove that Eq. (29) can be reduced to the dispersion relation of reactive two stream instability $1 - \omega_{be}^2/(\omega - kv_{be})^2 - \omega_{pe}^2/\omega^2 = 0$. Then, following similar reasoning as Ref. [88,94], taking wave trapping into account, an estimate of initial saturated wave amplitude $|\tilde{E}_{h0}|$ can be obtained (see Eq. (24) in main text). The subsequent wave-wave instabilities described by $\tilde{E}_{h1}$ could be further considered on this wave packet as shown below. This is how our model could self-consistently serve as one of the solutions to the problem.

In order to obtain the expression for $n_l$ in Eq. (29), we move on to treat the low frequency evolution of the bulk electrons (i.e., Eq. (16)). According to the assumption stated above, Eq. (16) can be reduced to

$$\gamma_e T_e \frac{\partial}{\partial x} \ln n_l = -e\boldsymbol{E}_l - \frac{m}{2}\frac{\partial}{\partial x}\overline{\boldsymbol{u}_h^2}. \tag{30}$$

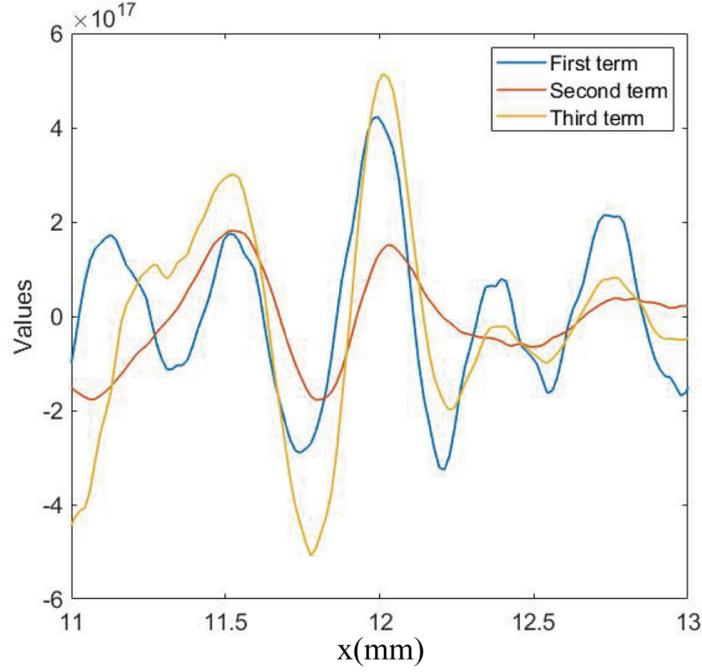

**Figure A1**: Balance of terms in Eq. (31) for Case 2. The data is taken from the particle simulation at $t = 128.4 \ ns$ and $y = 4.2 \ mm$. The time averaging is over the centered window of $15.1 ns$ (50 electron plasma periods). Since the instability is on ion time scale, the averaging window is of that order. We see that the first term (blue line) nearly balances the third term (yellow line).

Taking a spatial derivative and combining Eq. (18) with Eq. (30), we obtain:

$$\frac{\gamma_e T_e}{m}\frac{\partial^2}{\partial x^2}\ln n_l - \frac{e^2}{\epsilon_0 m}(n_{bl} + n_l - n_i) = -\frac{1}{2}\frac{\partial^2}{\partial x^2}\overline{\boldsymbol{u}_h^2}. \tag{31}$$

***This equation is central to the discussion below***, as different physical regimes, including the one newly discovered, correspond to the different relative ordering of the three terms. The physical meaning of Eq. (31) is that the ponderomotive force (right hand side) exerted on the plasma is balanced by the pressure force of the bulk electrons (first term) as well as the electric field caused by charge separation (second term).

The first two terms compare as $\frac{\gamma_e T_e \frac{\partial^2}{\partial x^2} \ln n_l}{\frac{e^2}{\epsilon_0}(n_{bl}+n_l-n_i)} \sim \gamma_e k^2 \lambda_{De}^2 \frac{n_l - n_0}{n_l}$ while the last two terms compare as

$$\frac{\frac{m}{2}\frac{\partial^2}{\partial x^2}\overline{u_h^2}}{\frac{e^2}{\epsilon_0}(n_{bl}+n_l-n_i)} \sim k^2 \lambda_{De}^2 \frac{\epsilon_0 |\widetilde{E}_h|^2}{4n_0 T_e} \frac{n_0}{n_l+n_{bl}-n_i} = \frac{1}{4}(k^2\lambda_{De}^2)^2 \left(\frac{e|\Phi_{h|}}{T_e}\right)^2 \frac{n_0}{n_l+n_{bl}-n_i}.$$

Given that $k^2 \lambda_{De}^2 \ll 1$, for the first two terms to be comparable, $n_l + n_{bl} - n_i$ must be small and $n_l - n_0$ must be relatively large. For the last two terms to be comparable, the time-averaged charge imbalance $n_l + n_{bl} - n_i$ should small or $\frac{\epsilon_0 |\widetilde{E}_h|^2}{4n_0 T_e}$ should be large.

There are different physical regimes that correspond to different assumptions regarding the ordering of non-dimensional parameters appearing above. In what follows, those regimes will be considered case-by-case.

### 1. Standing Wave Modulation Instability (SWMI)

The first case is to assume that the instability evolves sufficiently slowly, and the ions have time to respond to the ponderomotive force. Under this assumption, we then use charge neutrality condition $n_{bl} + n_l - n_i \approx 0$ for the initial phase of the instability.

The balance of terms of Eq. (31) in a simulation for this case is shown in Fig. A1. Integrating Eq. (31) from $x_0$ to $x$, where $x_0$ is the location where $n_l \sim n_0$, $\overline{u_h^2} \sim 0$, $\frac{\partial}{\partial x}\ln n_l \sim 0$, $\frac{\partial}{\partial x}\overline{u_h^2} \sim 0$, we obtain:

$$n_l \approx n_0 - \frac{mn_0}{2T_e}\overline{u_h^2}, \tag{32}$$

where we have used $\ln(n_l - n_0)/n_0 \approx (n_l - n_0)/n_0$. Note that such approximation is valid only at the onset of the instability. Let us return to Eq. (29). We now consider the slowly varying envelope of the high frequency field $E_h$. Expressing it as $E_h(x,t) = \frac{1}{2}[\widetilde{E}_h(x,t)e^{-i\omega_0 t} + \widetilde{E}_h^\star(x,t)e^{i\omega_0 t}]$ and doing likewise for $S$, we have

$$\left(2i\omega_0 \frac{\partial}{\partial t} + 2i\omega_0 \Gamma_e \frac{n_l}{n_0} + \omega_0^2 - \frac{n_l e^2}{\epsilon_0 m} + \frac{3T_e}{m}\frac{\partial^2}{\partial x^2}\right)\widetilde{E}_h = \widetilde{S}. \tag{33}$$

Both $\widetilde{E}_h$ and $\widetilde{S}$ are complex-valued functions that describe the evolution of the wave envelope. Using Eq. (32) and the approximation $\overline{u_h^2} \approx \epsilon_0 \overline{E_h^2}/n_0 m = \epsilon_0 |\widetilde{E}_h|^2/2n_0 m$ as well as $\frac{n_l}{n_0} \approx 1$ in the second term of (33), and taking $\omega_0 \approx \omega_{pe}$ (which is reasonable from the simulation results of Case 1 (see Fig. 5)), we obtain

$$\left(2i\omega_{pe}\frac{\partial}{\partial t} + 2i\omega_{pe}\Gamma_e + \frac{\omega_{pe}^2 \epsilon_0 |\widetilde{E}_h|^2}{4n_0 T_e} + \frac{3T_e}{m}\frac{\partial^2}{\partial x^2}\right)\widetilde{E}_h = \widetilde{S}. \tag{34}$$

Eq. (34) is the revised Zakharov first equation for the electric field envelope. In this equation, the envelope of $\widetilde{E}_h$ contains the contribution from both the initial packet and the perturbation. We can separate $\widetilde{E}_h$ as: $\widetilde{E}_h = \widetilde{E}_{h0} + \widetilde{E}_{h1}$ ($\widetilde{E}_{h1} \ll \widetilde{E}_{h0}$). Eq. (34) then gives rise to two equations for $\widetilde{E}_{h0}$ and for $\widetilde{E}_{h1}$:

$$\left(2i\omega_{pe}\frac{\partial}{\partial t} + 2i\omega_{pe}\Gamma_e + \frac{\omega_{pe}^2 \epsilon_0 |\widetilde{E}_{h0}|^2}{4n_0 T_e} + \frac{3T_e}{m}\frac{\partial^2}{\partial x^2}\right)\widetilde{E}_{h0} = \widetilde{S}, \tag{35}$$

$$\left(2i\omega_{pe}\frac{\partial}{\partial t} + 2i\omega_{pe}\Gamma_e + \frac{\omega_{pe}^2 \epsilon_0 |\widetilde{E}_{h0}|^2}{2n_0 T_e} + \frac{3T_e}{m}\frac{\partial^2}{\partial x^2}\right)\widetilde{E}_{h1} + \frac{\omega_{pe}^2 \epsilon_0 \widetilde{E}_{h0}^2}{4n_0 T_e}\widetilde{E}_{h1}^* = 0. \tag{36}$$

Note that in this Appendix didn't actually solve Eq. (35) directly. When comparing with simulations, the initial saturation level $|\widetilde{E}_{h0}|$ is estimated using Eq. (24) in the main text. Eq. (36) is our main focus below. The excitation source $\widetilde{S}$ is not of great interest here since our purpose is to obtain a purely growing solution of $\widetilde{E}_{h1}$. We assume that $\widetilde{S}$ varies very slowly in time. One needs to note that the terms such as $n_h u_h - \overline{n_h u_h}$ contributing to $S$ would disappear in the expression of $\widetilde{S}$, since they both describe slowly varying envelopes. The remaining terms in $\widetilde{S}$ are due to the electron beam. Physically, our procedure means that we assume an initial stationary wave with envelope $\widetilde{E}_{h0}$ already excited by the beam, with $\widetilde{E}_{h1}$ being a small perturbation to this wave envelope. Taking a complex conjugate to Eq. (36), the latter equation becomes:

$$\left(-2i\omega_{pe}\frac{\partial}{\partial t} - 2i\omega_{pe}\Gamma_e + \frac{\omega_{pe}^2 \epsilon_0 |\widetilde{E}_{h0}|^2}{2n_0 T_e} + \frac{3T_e}{m}\frac{\partial^2}{\partial x^2}\right)\widetilde{E}_{h1}^* + \frac{\omega_{pe}^2 \epsilon_0 \widetilde{E}_{h0}^{*2}}{4n_0 T_e}\widetilde{E}_{h1} = 0. \tag{37}$$

To obtain the instability threshold, we combine Eqs. (36) and (37). Assuming a perturbation of the form $\widetilde{E}_{h1} \sim e^{ikx + \gamma_{SWMI} t}$ and setting $\gamma_{SWMI} = 0$, we obtain:

$$-3\left(\frac{\epsilon_0 |\widetilde{E}_{h0}|^2}{4n_0 T_e}\right)^2 + 12k^2 \lambda_{De}^2 \frac{\epsilon_0 |\widetilde{E}_{h0}|^2}{4n_0 T_e} - 9k^4 \lambda_{De}^4 - \left(2\frac{\Gamma_e}{\omega_{pe}}\right)^2 = 0. \tag{38}$$

The physical meaning of this instability is self-localization of the wave envelope, which gives rise to the standing wave observed in our simulations. We therefore call such an instability Standing Wave Modulational Instability (SWMI). Without damping, the threshold is $\epsilon_0|\widetilde{E}_{h0}|^2/4n_0T_e > (k_0\lambda_{De})^2$. With accounting for damping only, in the case of very long waves, we need $\epsilon_0|\widetilde{E}_{h0}|^2/4n_0T_e > 2\Gamma_e/\omega_{pe}$ in order to obtain a positive value of the growth rate. Therefore, the real threshold obtained from this equation is (see also the main text):

$$\frac{\epsilon_0|\widetilde{E}_{h0}|^2}{4n_0T_e} = \max\left[\frac{2\Gamma_e}{\omega_{pe}}, (k_0\lambda_{De})^2\right]. \tag{39}$$

*Physically, it means that the electric field must be strong enough to overcome the plasma wave dispersion for the nonlinear self-localization to occur. At the same time, the damping must be small enough so that the wave could grow locally.* Note that the growth rate of this instability must be still on the ion time scale, namely, $\gamma_{SWMI} \lesssim \omega_{IAW}$, since $n_{bl} + n_l - n_i \approx 0$ is our pre-assumption. To illustrate the physics described by Eq. (36), we apply spatial Fourier transform to that equation and to its complex conjugate, add the two together, and neglect the time derivatives to obtain:

$$2\left(|\widetilde{E}_{h0}|^2\widetilde{E}_{h1}\right)(k) \sim$$
$$\int dk' \widetilde{E}_{h0}(k-k')\widetilde{E}_{h0}(k-k')\widetilde{E}_{h1}^*(k') + \int dk' \widetilde{E}_{h0}(k+k')\widetilde{E}_{h0}(k+k')\widetilde{E}_{h1}(k'). \tag{40}$$

Here $k' \ll k$. We can see that this equation could be understood as a four-wave coupling process $2L_0 \rightarrow L_{low} + L_{up}$. Note that this equation is only used to help qualitatively understand the physics as a four-wave interaction process. It is not employed in quantitative comparison with simulations.

The low-frequency ion dynamics can also be easily expressed by combining Eq. (15), Eq. (17), and Eq. (30):

$$\left(\frac{\partial^2}{\partial t^2} + \Gamma_i\frac{\partial}{\partial t} - \frac{\gamma_i T_i}{M}\frac{\partial^2}{\partial x^2}\right)\frac{n_i}{n_0} - \frac{T_e}{M}\frac{n_i}{n_l}\frac{\partial^2}{\partial x^2}\frac{n_l}{n_0} = \frac{\epsilon_0}{4Mn_0}\frac{\partial^2}{\partial x^2}|\widetilde{E}_h|^2, \tag{41}$$

where the same approach to damping has been used as in the electron case: $\Gamma_i = \Gamma_{IAW} + v_i/2$. Eq. (33), Eq. (31) and Eq. (41) *together constitute the modified Zakharov equations*.

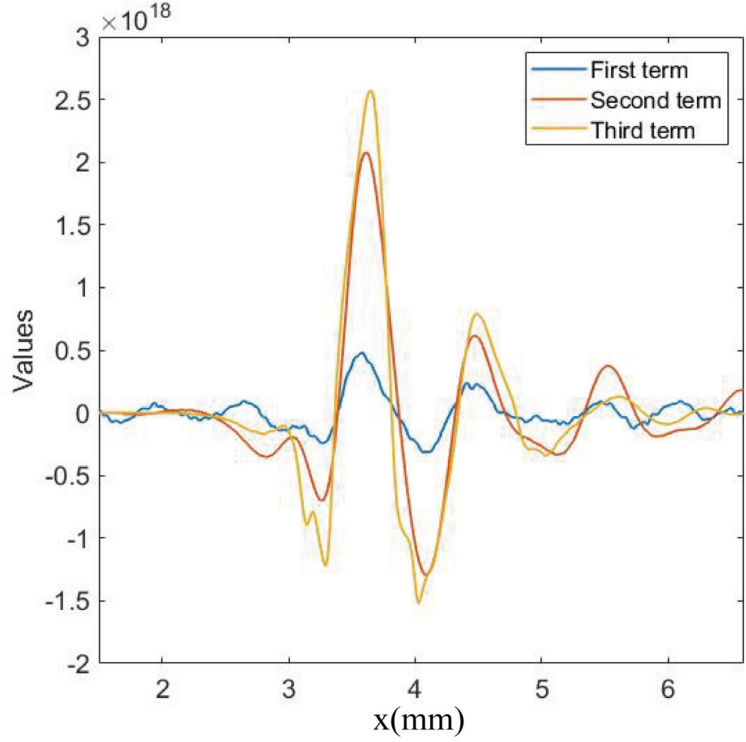

**Figure A2**: Balance of terms in Eq. (31) for the simulated Case 1. The data is taken from Case 1 at $t = 16.1\ ns$ and $y = 6.8\ mm$. The time average is taken over the window of $3.0ns$ (10 electron plasma periods, i.e., on electron time scale). The convergence with respect to time resolution had been verified. It is seen that the second term (red line) nearly balances the third term (yellow line).

## 2. Newly identified Electron Modulation Instability (EMI)

The second case is one when electrons evolve much faster than ions so that the ions don't have time to respond and could be regarded as a stationary background $n_i \approx n_0$. To the best of our knowledge, this case has not been studied by previous authors. The simulation Case 1 in the main paper corresponds to this case.

The balance of terms of Eq. (31) in our simulation for this case is illustrated in Fig. A2. The ordering of the two terms on the left-hand side of Eq. (31) is $\frac{e^2}{\epsilon_0}(n_{bl} + n_l - n_i)/(T_e \frac{\partial^2}{\partial x^2} \ln n_l) \sim \frac{1}{k^2 \lambda_{De}^2} \gg 1$. We therefore obtain approximately:

$$n_l \approx n_i - n_{bl} + \frac{m\epsilon_0}{2e^2}\frac{\partial^2}{\partial x^2}\overline{u_h^2}. \tag{42}$$

Note that such approximation is valid only at the onset of the instability, before ions have time to react. Applying the same approach as in the previous case, we substitute $\boldsymbol{E}_h(x,t) = \frac{1}{2}[\widetilde{\boldsymbol{E}}_h(x,t)e^{-i\omega_0 t} + \widetilde{\boldsymbol{E}}_h^*(x,t)e^{i\omega_0 t}]$ into Eq. (29) to obtain

$$\left(2i\omega_0 \frac{\partial}{\partial t} + 2i\omega_0 \Gamma_e \frac{n_l}{n_0} + \omega_0^2 - \frac{n_l e^2}{\epsilon_0 m} + \frac{3T_e}{m}\frac{\partial^2}{\partial x^2}\right)\widetilde{\boldsymbol{E}}_h = \widetilde{\boldsymbol{S}}. \tag{43}$$

Both $\widetilde{\boldsymbol{E}}_h$ and $\widetilde{\boldsymbol{S}}$ are complex-valued functions. Under the same approximations as used in the previous case, we obtain:

$$\omega_{pe}^2\left(2i\frac{1}{\omega_{pe}}\frac{\partial}{\partial t} + 2i\frac{\Gamma_e}{\omega_{pe}} - \left(\lambda_{De}^2\frac{\epsilon_0}{4n_0 T_e}\frac{\partial^2}{\partial x^2}|\widetilde{\boldsymbol{E}}_h|^2 - \frac{n_{bl}}{n_0}\right) + 3\lambda_{De}^2\frac{\partial^2}{\partial x^2}\right)\widetilde{\boldsymbol{E}}_h = \widetilde{\boldsymbol{S}}, \tag{44}$$

where $n_i \approx n_0$ is assumed since we are considering an instability whose growth time is short on the ion time scale. Eq. (44) is the revised Zakharov first equation for the electric field envelope. Suppose the beam is not very strong, namely $\frac{n_{bl}}{n_0} < 3k^2\lambda_{De}^2$. Upon expressing $\widetilde{\boldsymbol{E}}_h$ again as $\widetilde{\boldsymbol{E}}_h = \widetilde{\boldsymbol{E}}_{h0} + \widetilde{\boldsymbol{E}}_{h1}$ ($\widetilde{\boldsymbol{E}}_{h1} \ll \widetilde{\boldsymbol{E}}_{h0}$), Eq. (44) can be split into two equations for $\widetilde{\boldsymbol{E}}_{h0}$ and for $\widetilde{\boldsymbol{E}}_{h1}$:

$$\left(2i\omega_{pe}\frac{\partial}{\partial t} + 2i\omega_{pe}\Gamma_e - \frac{\epsilon_0}{4m_e n_0}\frac{\partial^2}{\partial x^2}|\widetilde{\boldsymbol{E}}_{h0}|^2 + \frac{n_{bl}}{n_0}\omega_{pe}^2 + \frac{3T_e}{m}\frac{\partial^2}{\partial x^2}\right)\widetilde{\boldsymbol{E}}_{h0} = \widetilde{\boldsymbol{S}}, \tag{45}$$

$$\left(2i\omega_{pe}\frac{\partial}{\partial t} + 2i\omega_{pe}\Gamma_e\right)\widetilde{\boldsymbol{E}}_{h1} - \frac{\epsilon_0}{4m_e n_0}\widetilde{\boldsymbol{E}}_{h1}\frac{\partial^2}{\partial x^2}|\widetilde{\boldsymbol{E}}_{h0}|^2 - \frac{\epsilon_0}{4m_e n_0}\widetilde{\boldsymbol{E}}_{h0}\frac{\partial^2}{\partial x^2}(\widetilde{\boldsymbol{E}}_{h0}^*\widetilde{\boldsymbol{E}}_{h1} +$$

$$\widetilde{\boldsymbol{E}}_{h0}\widetilde{\boldsymbol{E}}_{h1}^*) + \left(\frac{n_{bl}}{n_0}\omega_{pe}^2 + \frac{3T_e}{m}\frac{\partial^2}{\partial x^2}\right)\widetilde{\boldsymbol{E}}_{h1} = 0. \tag{46}$$

Note that here we also didn't actually solve Eq. (45) directly. When comparing with PIC simulations, the initial saturation level $|\widetilde{\boldsymbol{E}}_{h0}|$ is estimated using Eq. (24) in the main text. *Again, similarly to the previous section, Eq. (46) could be regarded as a rapid four-wave process.* Combining Eq. (46) with its complex conjugate, assuming the perturbation to obey the form $\widetilde{\boldsymbol{E}}_{h1} \sim e^{ikx + \gamma_{EMI}t}$ and writing $\frac{\partial^2}{\partial x^2} = -k^2$, we approximately find the growth rate of the instability from

$$-3\left(k^2\lambda_{De}^2\frac{\epsilon_0|\widetilde{\boldsymbol{E}}_{h0}|^2}{4n_0 T_e}\right)^2 + 4\left(3k^2\lambda_{De}^2 - \frac{n_{bl}}{n_0}\right)\left(k^2\lambda_{De}^2\frac{\epsilon_0|\widetilde{\boldsymbol{E}}_{h0}|^2}{4n_0 T_e}\right) - \left(3k^2\lambda_{De}^2 - \frac{n_{bl}}{n_0}\right)^2 -$$

$$\left(2\frac{\Gamma_e}{\omega_{pe}} + 2\frac{\gamma_{EMI}}{\omega_{pe}}\right)^2 = 0. \tag{47}$$

This is a case of a strong *electron-only modulational instability.* We therefore call it Electron Modulational Instability (EMI). The expression for the threshold takes the following form:

$$\frac{\epsilon_0 |\widetilde{E}_{h0}|^2}{4 n_0 T_e} = \max\left[1 - \frac{n_b}{3 n_0} \frac{1}{k^2 \lambda_{De}^2}, 2 \frac{\Gamma_e}{\omega_{pe}} \frac{1}{k^2 \lambda_{De}^2}\right]. \tag{48}$$

*Physically, it means that the electric field must be strong enough to modify the electron dynamics before ions could respond. At the same time, the damping must be small enough so that the wave could grow locally.* Here we take $n_{bl} \approx n_b$. The beam becomes dominant when $n_{bl}/n_0 > 3 k^2 \lambda_{De}^2$; in this case, Eq. (44) becomes:

$$\omega_{pe}^2 \left(2i \frac{1}{\omega_{pe}} \frac{\partial}{\partial t} + 2i \frac{\Gamma_e}{\omega_{pe}} + \frac{n_{bl}}{n_0} - \lambda_{De}^2 \frac{\epsilon_0}{4 n_0 T_e} \frac{\partial^2}{\partial x^2} |\widetilde{E}_h|^2 + 3 \lambda_{De}^2 \frac{\partial^2}{\partial x^2}\right) \widetilde{E}_h = \widetilde{S}. \tag{49}$$

For perturbed field $\widetilde{E}_{h1}$, it would be written as:

$$\left(2i \frac{1}{\omega_{pe}} \frac{\partial}{\partial t} + 2i \frac{\Gamma_e}{\omega_{pe}} + \frac{n_{bl}}{n_0} - \lambda_{De}^2 \frac{\epsilon_0}{4 n_0 T_e} \frac{\partial^2}{\partial x^2} |\widetilde{E}_{h0}|^2 + 3 \lambda_{De}^2 \frac{\partial^2}{\partial x^2}\right) \widetilde{E}_{h1} = 0. \tag{50}$$

Applying a similar approach to $\widetilde{E}_{h1}$, we get the growth rate:

$$2 \frac{\gamma_{EMI}}{\omega_{pe}} = \frac{n_{bl}}{n_0} + k^2 \lambda_{De}^2 \frac{\epsilon_0}{4 n_0 T_e} |\widetilde{E}_{h0}|^2 - 3 k^2 \lambda_{De}^2 - 2 \frac{\Gamma_e}{\omega_{pe}}. \tag{51}$$

The instability threshold for this case is

$$\frac{E_b}{T_e} = \left(\frac{2}{3} \frac{n_{bl}}{n_0} + \frac{3}{8} \left(\frac{n_b}{n_0}\right)^{4/3} - \frac{4}{3} \frac{\Gamma_e}{\omega_{pe}}\right)^{-1}. \tag{52}$$

*Physically, these results indicate that the wave envelope localization occurs faster than the ion time scale. In this case, the growth occurs before ions could have time to respond, which is exactly what takes place in Case 1.* This process results in charge separation that balances the ponderomotive force, which therefore gives a different physical picture from the traditional Langmuir collapse. The charge separation is clearly seen from the plots shown in Fig. A3. We also ran a test with ions fixed as a stationary background and verified that the wave would still grow significantly. Please note that when the ions start to move, Eq. (44) needs to be coupled with an ion equation. Here we only consider the initial stage of the instability where the ions do not move. The equation for ions in this case is still Eq. (41).

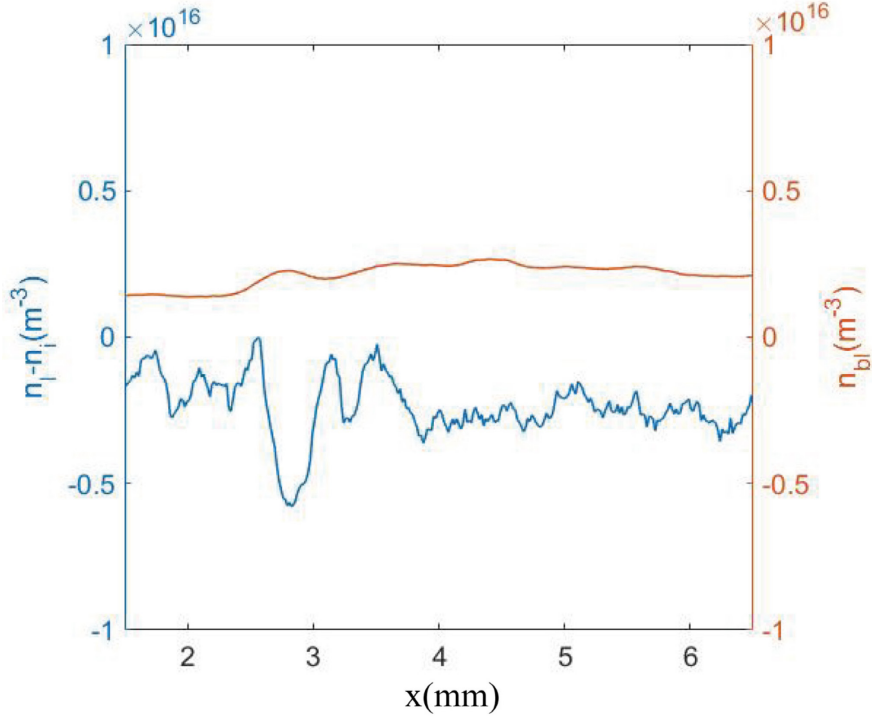

**Figure A3**: Profiles of $n_l - n_i$ and $n_{bl}$. The data is taken from Case 1 at $t = 45.2$ ns and $y = 6.8$mm, when the Langmuir wave envelope has grown by a large factor. The centered time-averaging window is $3.0 ns$ (10 electron plasma periods, i.e., on the electron time scale).

## 3. "Classical" case

For the third case we assume an initial stationary wave envelope, but in addition that the quasi-neutrality condition $n_l - n_i = 0$ is satisfied at any time. We also neglect the effect of the driving beam. This case is the one most studied by previous authors.

Taking $n_i = n_l$ and a small ion density modulation, Eq. (34) and Eq. (41) could be reduced to the traditional Zakharov equations used in the past [76]:

$$\left(2i\omega_{pe}\frac{\partial}{\partial t} + 2i\omega_{pe}\Gamma_e + \frac{3T_e}{m}\frac{\partial^2}{\partial x^2}\right)\widetilde{E}_h = \omega_{pe}^2 \frac{\delta n_i}{n_0}\widetilde{E}_h, \tag{53}$$

$$\left(\frac{\partial^2}{\partial t^2} + \Gamma_i\frac{\partial}{\partial t} - c_s^2\frac{\partial^2}{\partial x^2}\right)\frac{n_i}{n_0} = \frac{\epsilon_0}{4Mn_0}\frac{\partial^2}{\partial x^2}\left|\widetilde{E}_h\right|^2. \tag{54}$$

Note that we already neglected the source term. One may want to follow the approach of [2] for more detailed analysis. Here we only provide a brief discussion. Expressing $\widetilde{E}_h$ once again as $\widetilde{E}_h = \widetilde{E}_{h0} + \widetilde{E}_{h1}$ ($\widetilde{E}_{h1} \ll \widetilde{E}_{h0}$), where $\widetilde{E}_{h0}$ is the initial wave envelope giving the initial ion density depletion, we obtain:

$$\left(2i\omega_{pe}\frac{\partial}{\partial t} + 2i\omega_{pe}\Gamma_e + \frac{3T_e}{m}\frac{\partial^2}{\partial x^2}\right)\widetilde{E}_{h1} = \omega_{pe}^2 \frac{\delta n_i}{n_0}\widetilde{E}_{h0}, \tag{55}$$

$$\left(\frac{\partial^2}{\partial t^2} + \Gamma_i \frac{\partial}{\partial t} - c_s^2 \frac{\partial^2}{\partial x^2}\right)\frac{\delta n_i}{n_0} = \frac{\epsilon_0}{4Mn_0}\frac{\partial^2}{\partial x^2}\left(\widetilde{\bm{E}}_{h0}^*\widetilde{\bm{E}}_{h1} + \widetilde{\bm{E}}_{h0}\widetilde{\bm{E}}_{h1}^*\right). \tag{56}$$

Note that in this case, the ion density already had an initial depletion caused by $\left|\widetilde{\bm{E}}_{h0}\right|^2$, which is why the equations take a form different from what has been presented above. The density perturbation $\delta n_i$ is small compared to that depletion. Now we take $\widetilde{\bm{E}}_{h1} \sim e^{ikx-i\omega t}$, $\delta n_i \sim e^{ikx-i\omega t}$, and combine the two equations together. Under the assumption that $\widetilde{\bm{E}}_{h0}$ is nearly uniform, we substitute (55) into (56) to obtain:

$$(-\omega^2 - i\omega\Gamma_i + c_s^2 k^2)\frac{\delta n_i}{n_0} =$$

$$-\omega_{pe}^2 \frac{\epsilon_0 k^2}{4Mn_0}\left|\widetilde{\bm{E}}_{h0}\right|^2 \left(\frac{1}{2\omega_{pe}\omega + 2i\omega_{pe}\Gamma_e - \frac{3T_e}{m}k^2} + \frac{1}{2\omega_{pe}\omega^* - 2i\omega_{pe}\Gamma_e - \frac{3T_e}{m}k^2}\right)\frac{\delta n_i}{n_0}. \tag{57}$$

After tedious but straightforward derivations and using a plane wave assumption, we can prove that Eq. (57) can be reduced to exactly the same form as the traditional wave-coupling equations given by Nishikawa [58] (Eq. (31) therein). This implies that the traditional Zakharov equations are indeed an expansion of the traditional wave coupling equations. The rest of the analysis is the same as what Nishikawa did in his famous paper. The threshold for PDI in the main paper takes the form of Eq. (63) in Nishikawa's paper. However, setting $n_i = n_l$ is a too strong condition for our simulations since we also have the $n_{bl}$ beam component. The plane wave assumption employed by previous authors [58,76,86] is also an over-simplification in our case. We believe that SWMI gives a more reasonable threshold for the wave self-localization observed in the simulations.

### Discussion in the context of previous work

According to Eq. (29), the high-frequency evolution of the electric field in depends on the electron dynamics only. This is a very important concept. It means that it is low frequency electron dynamics instead of ion dynamics that is responsible for the wave self-localization. Of course, the electron and ion dynamics are coupled, and the charge quasi-neutrality condition approximately holds in many cases. However, as we have shown with EMI, the charge neutrality condition could break down if the field is strong enough. Another way to solve Eq. (37), (46), (53) is to assume a pre-exist field profile and density profile and use Hermite polynomial to express the solutions, similar to a 1D harmonic oscillator problem [84,95]. This method, despite nicely done, was also not fully self-consistent and was not convenient for obtaining a concise expression for threshold and growth rate (which is why we didn't use it). One should always remember that the analytical theory we provided here is valid *only* at the very beginning of the instability in our simulations.

If one neglects the right-hand side, Eq. (34) looks very similar to the result of Bellan, namely Eq. (15.133) in his book [75]. The difference is because Bellan used $n_l \approx n_i$ whereas here we adopted $n_l \approx n_i - n_{bl}$. Comparing with the work by Papadopoulos [86], our first case and second case do not assume $n_l \approx n_i$ and we also do not assume spatially homogeneous plane wave (i.e. we retained the spatial derivative term in Eq. (34) and Eq. (44)). We also notice that the electron density perturbation $\delta n_e$ may not be small in practical cases (due to the large amplitude of the Langmuir wave). Papadopoulos' 1974 work indeed follows the same approach as that of Nishikawa [58], which is a simplification of the traditional Zakharov model. Careful readers may also inquire about the difference between our approach and that due to Morales and Lee from 1976 [84]. Their approach is brilliant in that the authors not only considered the non-uniformity of the wave packet, but also the coupling of the high-frequency equation and the low-frequency equation. Moreover, the authors considered the trapping of modes in the density depletions, which has been a non-linear approach. However, Morales and Lee introduced the same assumption as the one behind the first case in our work. As seen from Eq. (31), it is not automatically valid in general. The ponderomotive force should act upon both electrons and ions. Since electrons respond faster, this may result in charge separation and a low frequency electric field that balances the ponderomotive force. It should be a three-term balance rather than one between two terms only.

**Evolution as transition between different regimes**

In the parameter space, different regimes have been mapped according to the linearized solutions the respective instabilities of the Langmuir wave envelope. In simulations, for each case under the strong turbulent regime (refer to corresponding figures in the main paper), we can visualize the evolution by imagining that the point on the map would move down after the saturation of modulational instability in each burst due to the increase of bulk electron temperature. Eventually, a point initially in a EMI regime could move to SWMI regime or PDI regime. However, the "speed" of this downward movement and the time that a point could stay in EMI regime, as mentioned in Sec. 3.5.2, depends on various factors. The actual simulation condition could be much more complicated than the three limiting regimes discussed in this Appendix. *We therefore emphasize that these theoretical developments mainly serve to help reader understand the physics behind simulations*.

**Summary**

In this Appendix, we re-derived the wave coupling equations starting from a two-fluid model and obtained a set of revised Zakharov equations. We found three different regimes corresponding to how the ponderomotive force-balance equation (31) is reduced. We found a threshold for the Langmuir wave packet self-localization under the conditions of SWMI, which determines the boundary of the Strong Turbulent Regime. A very fast instability called EMI has been demonstrated for the first time. It does not involve ion dynamics at the initial stage of its growth. We have also accounted for the effects of damping and non-uniform wave envelope.

Finally, we recall the assumptions made in arriving at the revised Zakharov equations presented in this work.

1. Fluid approximation $k_0^2 \lambda_{De}^2 \ll 1$ for bulk components. The wave-bulk particle resonance is neglected since we are considering wave-wave nonlinear interactions. That is not to say that wave-particle interactions are not important. Wave-particle interaction plays an important role in determining the initial saturated electric field as well as the shape of the wave envelope. Here we assume that the wave envelope has already been generated by the beam-plasma interaction.

2. 1D approximation.
3. Terms containing $u_l$ are small compared with other terms.
4. $n_h/n_l$ is not very large. Not very large means that $\ln(1 + n_h/n_l) \approx n_h/n_l$.
5. Other assumptions mentioned in the text.

## Appendix 2: Qualitative justification of setting $\gamma_e = 3$ for electrons in high-frequency motions.

In this section, we provide a simple argument for choosing the value of adiabatic index for electrons on the high frequency component (short time scale), $\omega \gg kv_{the}$. The usual reasoning is that fluid closure with $\gamma_e = 3$ provides the correct thermal dispersion for the Langmuir wave, consistent with the kinetic theory. However, it would be instructive to find $\gamma$ directly as $\partial p/\partial \rho$. To this end, we consider a monochromatic wave in the reference frame moving with the phase velocity $u = \omega/k$ in which the profile of the electrostatic potential $\Phi$, with an amplitude $\Phi_{max}$, is time stationary. We consider the Vlasov equation for the electrons under the conditions $u \gg \sqrt{T_e/m}$ and $u \gg \sqrt{e\Phi_{max}/m}$. The first one is the applicability condition $k\lambda_D \ll 1$ for the fluid approximation while the second one, $k\lambda_D\sqrt{e\Phi_{max}/T_e} \ll 1$, allows the kinetic equation to be linearized because the trajectories of the thermal electrons are weakly perturbed. We note that these two conditions are applicable *only* prior to the nonlinear stage of wave-wave instabilities. The linearized kinetic equation for the perturbed velocity distribution $f = f(v, x)$ of electrons is

$$\frac{\partial f}{\partial x} + \frac{e}{m}\frac{d\Phi}{dx}\frac{\partial f_0}{\partial v} = 0, \tag{58}$$

where $f_0 = f_0(v - u)$ is the equilibrium distribution localized at $|v - u| \ll u$. The solution is

$$f = \frac{e\Phi}{m}\frac{1}{v}\frac{df_0}{dv}. \tag{59}$$

The perturbations of mass density and pressure moments are then evaluated as follows:

$$\delta\rho = e\Phi \int dw \frac{1}{u+w}\frac{df_0}{dw} \approx \frac{e\Phi}{u^2}\rho_0, \tag{60}$$

$$\delta p = e\Phi \int dw \frac{w^2}{u+w}\frac{df_0}{dw} \approx -\frac{e\Phi}{u^2}\int w^3 \frac{df_0}{dw}dw = 3\frac{e\Phi}{u^2}p_0. \tag{61}$$

Therefore, within the linear approximation $\gamma = \frac{\partial p/\partial \Phi}{\partial \rho/\partial \Phi} = 3$ for an arbitrary shape of the equilibrium velocity distribution $f_0(v)$. The assumption $k\lambda_D\sqrt{e\Phi_{max}/T_e} \ll 1$ is not overly restrictive even if $e\Phi_{max}/T_e$ is not small and in our simulations at the initial stage of the instability, $k\lambda_D\sqrt{e\Phi_{max}/T_e}$ remains below 0.15. Still, the adiabatic index can be evaluated numerically without linearization because the velocity distribution can be expressed as a function of the energy integral $\frac{1}{2}mv^2 - e\Phi$. The adiabatic index $\gamma$ can then be calculated as a function of position $x$ within the wave period by evaluating the moments numerically and then carrying out a numerical differentiation with respect to $\Phi$. For example, the mass density is calculated as

$$\rho(x) = m \int dv \; f_0\left(\sqrt{v^2 - \frac{2e\Phi}{m}}\right), \tag{62}$$

and the second moment is an average of $(v - \langle v \rangle)^2$, where $\langle v \rangle$ is the first moment, equal to $u = \omega/k$ in the linear case.

Fig. A4 shows the maps of the $\gamma$ value calculated numerically for a Maxwellian $f_0$ as a function of the dimensionless parameters $e\Phi_{max}/2T_e$ and $k\lambda_D$. It is seen that within the given range of arguments, the adiabatic index deviates from 3 by within 10 percent. We note that it is always above 3 and the largest deviation occurs at the trough of the potential profile. In the end, we believe that setting $\gamma = 3$ is indeed a good assumption within the fluid model.

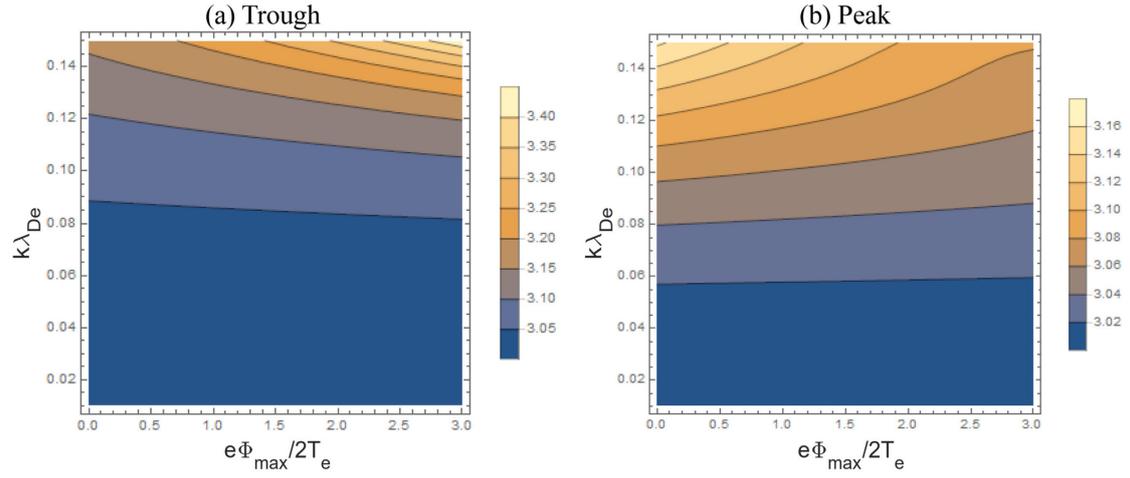

**Figure A4**: Adiabatic index for electrons in a sinusoidal profile of electrostatic potential vs $e\Phi_{max}/2T_e$ and $k\lambda_D$ (a) at the trough of the potential energy and (b) at the peak position.

# Appendix 3: Parameters of all the simulation cases

**Table A1**: All the 57 simulation cases with parameters: ratio of the beam density over plasma density $n_b/n_{p0}$, the initial energy of beam electrons $E_{b0}$, the initial temperature of bulk electrons $T_{e0}$, and system length in $x$ direction, $L_x$.

| Case | $n_b/n_{p0}$ | $E_{b0}(eV)$ | $T_{e0}(eV)$ | $L_x(mm)$ |
|---|---|---|---|---|
| 1 | 0.015 | 30 | 0.2 | 32 |
| 2 | 0.00068 | 45 | 0.2 | 32 |
| 3 | 0.015 | 30 | 2.0 | 32 |
| 4 | 0.0015 | 25 | 0.5 | 32 |
| 5 | 0.015 | 80 | 2.0 | 64 |
| 6 | 0.015 | 30 | 0.5 | 32 |
| 7 | 0.015 | 1 | 0.2 | 16 |
| 8 | 0.015 | 5 | 0.2 | 16 |
| 9 | 0.015 | 15 | 0.2 | 16 |
| 10 | 0.015 | 50 | 0.2 | 32 |
| 11 | 0.015 | 100 | 0.2 | 64 |
| 12 | 0.015 | 50 | 0.5 | 32 |
| 13 | 0.015 | 100 | 0.5 | 64 |
| 14 | 0.015 | 5 | 0.6 | 16 |
| 15 | 0.015 | 10 | 0.6 | 16 |
| 16 | 0.015 | 10 | 0.7 | 16 |
| 17 | 0.015 | 20 | 0.7 | 16 |
| 18 | 0.015 | 10 | 1.0 | 16 |
| 19 | 0.015 | 30 | 1.0 | 32 |
| 20 | 0.015 | 50 | 1.0 | 32 |
| 21 | 0.015 | 90 | 1.0 | 64 |
| 22 | 0.015 | 13 | 1.5 | 16 |
| 23 | 0.015 | 18 | 1.5 | 16 |
| 24 | 0.015 | 10 | 2.0 | 16 |
| 25 | 0.015 | 20 | 2.0 | 16 |
| 26 | 0.015 | 45 | 2.0 | 32 |
| 27 | 0.00068 | 5 | 0.2 | 16 |
| 28 | 0.00068 | 15 | 0.2 | 16 |
| 29 | 0.00068 | 25 | 0.2 | 32 |
| 30 | 0.00068 | 180 | 0.2 | 64 |
| 31 | 0.0015 | 10 | 0.5 | 16 |
| 32 | 0.0015 | 45 | 0.5 | 32 |
| 33 | 0.0015 | 60 | 0.5 | 32 |
| 34 | 0.003 | 40 | 0.2 | 32 |
| 35 | 0.003 | 100 | 0.2 | 64 |
| 36 | 0.003 | 10 | 0.5 | 16 |
| 37 | 0.003 | 13 | 0.5 | 16 |
| 38 | 0.003 | 15 | 0.5 | 16 |
| 39 | 0.003 | 25 | 0.5 | 32 |
| 40 | 0.003 | 45 | 0.5 | 32 |
| 41 | 0.0075 | 5 | 0.5 | 16 |
| 42 | 0.0075 | 10 | 0.5 | 16 |
| 43 | 0.0075 | 20 | 0.5 | 16 |
| 44 | 0.0075 | 30 | 0.5 | 32 |
| 45 | 0.0075 | 40 | 0.5 | 32 |
| 46 | 0.0075 | 130 | 0.5 | 64 |
| 47 | 0.03 | 10 | 0.5 | 16 |
| 48 | 0.03 | 15 | 0.5 | 16 |
| 49 | 0.03 | 50 | 1.0 | 32 |
| 50 | 0.03 | 10 | 2.0 | 16 |
| 51 | 0.03 | 15 | 2.0 | 16 |
| 52 | 0.03 | 25 | 2.0 | 32 |
| 53 | 0.03 | 40 | 2.0 | 32 |
| 54 | 0.03 | 70 | 2.0 | 64 |
| 55 | 0.07 | 7 | 1.0 | 16 |
| 56 | 0.07 | 13 | 1.0 | 16 |
| 57 | 0.07 | 30 | 1.0 | 32 |

## Appendix 4: Discussion of kappa distribution and energy spectrum

### Electron kappa distribution function

The kappa distribution function observed in Case 1 is truncated at energy corresponding to the wall potential because the particles with sufficient energy can overcome the retarding electric field in the anode sheath and get lost. To better study a formation of the kappa distribution function, we performed other two simulations with the same parameters as Case 1 and Case 3, but we collected and recorded the energetic bulk electrons lost at the anode and used the data to plot EVDF. The results are shown in Fig. 12. As we can see, both the EVDFs simulated in Case 1 and Case 3 show a high energy tail, which could be described by the kappa distribution and is consistent with previous work [26]. The difference between these two cases can also be clearly seen: strong electron heating occurs for Case 1 but not for Case 3. This is another evidence that bulk electron temperature could strongly affect the properties of wave heating. The physical reason could be that in strong turbulence, the wave-vector that contains most energy significantly increases (see Fig. 5 for Case 1), allowing the waves to transfer energy directly with bulk electrons ($k \sim \omega_{pe}/v$).

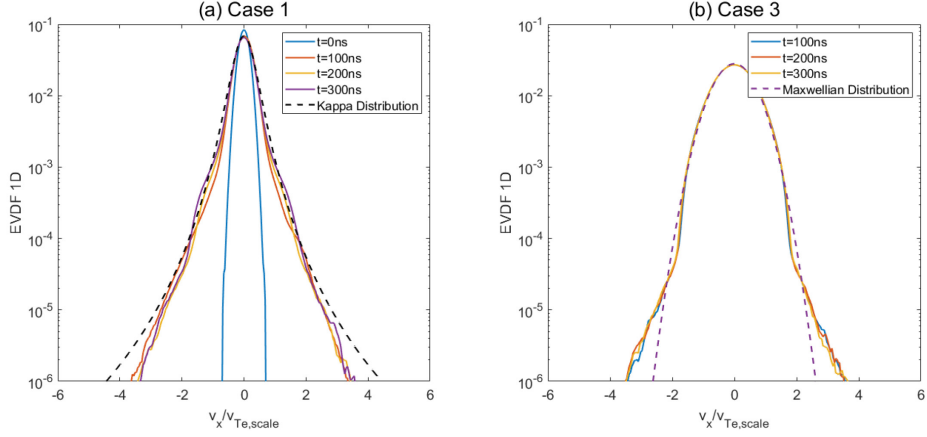

**Figure A5**: (a) and (b) show the 1D EVDF $f(v_x)$ with energetic bulk electrons collected for Case 1 at $t = 0ns, 100ns, 200ns, 300ns$ and Case 3 at $t = 100ns, 200ns, 300ns$. The black dashed line in (a) shows the kappa distribution with coefficients $\kappa = 1.6$ and $T = 1.0eV$. Here, the calculation of EVDF used all the particles in the simulation domain plus the collected at anode energetic particles. All EVDFs are normalized to unity by the integration of EVDF. EVDF in Case 1 resembles a kappa distribution and there is a stronger wave heating than in Case 3.

## Energy spectrum

Wave energy spectrum is a frequently used statistical method describing the energy cascade in turbulence [74]. A very interesting finding is the $k^{-5}$ power spectrum observed below the pump scale ($k_{pump} \approx \omega_{pe}/v_b$) when SWMI or EMI and associated strong wave-wave nonlinear interaction are occurring in the system. We know from wave-particle resonance condition that the waves with lower wave-vectors interact with super-thermal electrons ( $k \sim \omega_{pe}/v_{superthermal}$, where $v_{superthermal} > v_b$ ). According to the classical Strong Langmuir Turbulence theory, such a -5 spectrum could be explained by the interaction between super-thermal electrons and high-frequency waves [96,97], where the authors obtained a -4.5 spectrum, nearly approaching the -5 spectrum in the simulations.

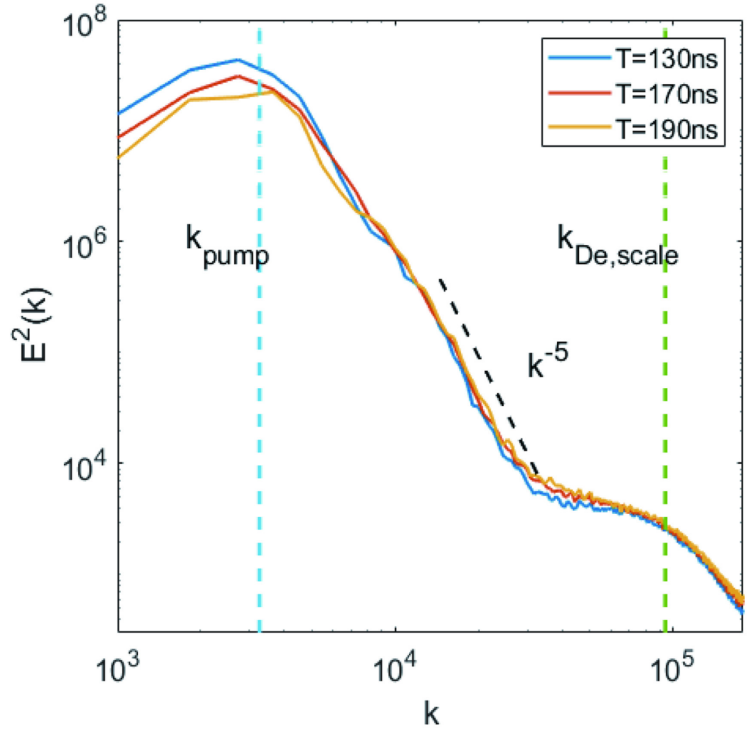

**Figure A6**: Energy spectrum plot for Case 30 in SWMI regime at $t = 130ns, 150ns, 170ns$. A $-5$ spectrum indicating presence of strong turbulence is observed. The pump length and the scaled Debye length are also shown by the vertical dashed lines.

Fig. A6 for Case 30 in the SWMI regime shows that the $k^{-5}$ spectrum is frequently observed for the cases in strong turbulent regime. We also noticed that for the wave number between $k = k_{pump}$ and $k \approx 10^4$, the power index indeed deviates a little bit from the $k^{-5}$ spectrum for this case. The exact physical reason for this is not fully understood yet. Therefore, we show that the strong Langmuir turbulence manifests new fascinating results that hasn't been fully understood. The exact theoretical analysis is left for future studies.